\documentclass{aa}

\usepackage{txfonts}
\usepackage{graphicx}
\usepackage{enumerate}
\usepackage{mathtools}
\makeatletter
\usepackage{subcaption}
\usepackage{aas_macros}
\usepackage{epstopdf}

\newcommand{\Rmnum}[1]{\expandafter\@slowromancap\romannumeral #1@}

\makeatother

\begin{document}

\title{G-virial: Gravity-based structure analysis of molecular
clouds\thanks{Available at \url{http://gxli.github.io/G-virial/.}} } 
 
\author{Guang-Xing Li \inst{1} \and  Friedrich
 Wyrowski\inst{1} \and Karl Menten\inst{1} \and Tom
 Megeath\inst{2} \and Xun Shi\inst{3}}

 \institute{Max-Planck Institut f\"ur Radioastronomie, Auf dem H\"ugel, 69, 
 53121 Bonn, Germany \and University of Toledo, Ritter Astrophysical
 Observatory, Department of Physics and Astronomy, Toledo OH 43606
 \and Max-Planck-Institut f\"ur Astrophysik, Karl-Schwarzschild-Stra{\ss}e 1,
 85740 Garching bei M\"unchen, Germany
 }
\offprints{Guang-Xing Li, \email{gxli@mpifr-bonn.mpg.de, gxli@usm.lmu.de}}
\abstract{We present the G-virial method which aims to quantify (1) the importance
of gravity in molecular clouds in the position-position-velocity (PPV) space, and (2) properties of the gas condensations in molecular clouds.
 Different from previous approaches that calculate the virial parameter for
different regions, our new method takes gravitational interactions between
all the voxels in 3D PPV data cubes into
account, and generates maps of the importance of gravity. This map can be
combined with the original data cube to derive relations such as the
mass-radius relation. Our method is important for several reasons.
First, it offers the the ability to quantify the centrally condensed structures
in the 3D PPV data cubes, and enables us to compare them in an
uniform framework.
Second, it allows us to understand the importance of gravity at different
locations in the data cube, and provides a {\it \emph{global}} picture of gravity in
clouds.
Third, it offers a  robust approach to decomposing the data into different
regions which are gravitationally coherent.
 To demonstrate the application of our method we identified regions from the
 Perseus and Ophiuchus molecular clouds, and analyzed their properties. We found an increase in the importance
 of gravity towards the centers of the individual molecular condensations.
We also quantified the properties of the regions in terms of mass-radius and
mass-velocity relations. Through evaluating the virial parameters based on
the \texttt{G-virial}, we  found that all our regions are almost
gravitationally bound. Cluster-forming regions appear are more centrally
condensed.}

\maketitle

\section{Introduction}
Star formation takes place in the dense and shielded parts of the 
interstellar medium.
Observations show that the interstellar medium exhibits
complicated, irregular, and filamentary structures
\citep{1979ApJS...41...87S,2000prpl.conf...97W,2008ApJ...680..428G,2010A&A...518L.103M}.
Theoretically, such structures are created by various physical processes
such as turbulence, gravity, magnetic field, and radiation.

To understand star formation it is necessary to understand how
various physical processes affect it. Gravity is a long-range interaction, and plays important
roles in most astrophysical processes at
 a multiple of physical scales. This is  particularly true for turbulent
 molecular clouds \citep[e.g.,][]{2009ApJ...699.1092H,2013arXiv1308.5679K}.
 However, characterizing the role of gravity on a cloud at various physical
 scales is not straightforward.

In observations, the structure of molecular gas can be traced by spectral lines
in the 3D PPV (position-position-velocity) space where molecular gas
exhibits complicated structures. However, given the rich information obtained
observationally, few constraints on the role of gravity in the clouds have been
obtained so far.
One major difficulty is to properly quantify the irregular structure of the gas.
The virial parameter $\alpha_{\rm vir}=5 \sigma^2 R/G M$ \footnote{Here $\sigma$ is the
velocity dispersion and $M$ is the mass \citep{1992ApJ...395..140B}.  The
surface terms and the magnetic terms are neglected. See 
\citet{2006MNRAS.372..443B} for a thorough discussion.}  is commonly used to
quantify the importance of gravity, and to calculate it we must define a region in which $\sigma$ and $M$ can be
evaluated.\footnote{Such regions have been defined with the \texttt{clumpfind}
\citep{1994ApJ...428..693W} and the \texttt{dendrogram} \citep{2008ApJ...679.1338R} programs in the
past.} As a result, the virial parameter
is only suitable for the cases where the structures are well defined. 
The morphology of the molecular interstellar medium is generally so complicated
that in many cases it is difficult to separate individual objects from a
continuous distribution of material. This has been further complicated by the
fact that we observe the structures of the sky plane,
and the structures identified from the observations are
biased by various projection effects \citep{2000ApJ...532..353P,
2006ApJ...638..797D, 2010ApJ...712.1049S, 2013ApJ...777..173B}.

 Another difficulty with the virial parameter is that it is {\it \emph{localized}}. 
 To define the virial parameter we need to
 define the region, and by evaluating the virial parameter of that region, we automatically neglect the gravitational
 interaction between the region and its surroundings. As a result, the virial
 parameter can be used to quantify the importance of gravity for the individual
 structures, but it cannot  be used to understand the importance of
 gravity for a cloud { as a whole}. Gravity is a {\it \emph{long-range}} interaction,
 and to understand its importance, we need to understand how it
 works not it one region on one particular physical scale, but on a multiple of
 physical scales.

 One case where such an understanding is required is the cluster-forming
 region. In observations, such regions are found to be centrally condensed in
 both the position-position space and the PPV space, and physically we expect that gravity
 is important at the centers of the cluster-forming regions and should become
 less important if we move from the centers to the outskirts. This
 stratification, which is of crucial importance for understanding stellar
 clustering, awaits to be quantified.

In this paper we introduce a new method called  \texttt{G-virial}
(avaliable at https://gxli.github.com/G-virial) to quantify the
importance of gravity in a variety of situations where mass is traced in the 3D PPV space.  Instead of dividing the molecular cloud into regions based on
  iso-intensity contours in the 3D data cube
  \citep[e.g.,][]{1994ApJ...428..693W}, the method takes the gravitational interactions between all the voxels in the
  3D PPV data cube into account, and generates maps of the importance of
  gravity in 3D.
  As a result, it provides constraints on the {\it \emph{global}} importance of
  gravity. This is complementary to methods that calculate the
  self-gravity such as the virial parameter.  
  Such a map of the importance of gravity also enables us to identify
  gravitationally coherent regions in molecular clouds and to quantify the
  structures of the regions.

 In this paper, we  introduce the   \texttt{G-virial}  method (section
 \ref{sec:formulation} -- \ref{sec:justification}), and explore its usage in quantifying
 the importance of gravity and the structures of the molecular condensations
 seen in simulations (section \ref{sec:justification}) and observations
 (section \ref{sec:appl}).
 We conclude in section \ref{sec:conclu}.

\section{Problem formulation }\label{sec:formulation}

Observationally, molecular
gas can be mapped by rotational transitions of molecules such as CO. Through
proper modeling of the emission, it is possible to construct the distribution
of the molecular gas in the form of a 3D PPV data cube
which covers a continuous $(x,y,v)$ space
\begin{equation}
  m(x,y,v)=\widetilde{\rho}(x,y,v)\;{\rm d}x\; {\rm d} y \;{\rm d}v\;,
\end{equation}
where  $m$ represents the mass, $x$ and $y$ represent the spatial dimensions, and
$v$ represents the velocity dimension.  Here $\widetilde{\rho}$ has a dimension
of $M\; L^{-3} T$, and represents the amount of mass per unit area per unit
velocity. The distribution of $\widetilde{\rho}$ in the $v$ dimension represents
the distribution of gas at different velocities along the same line of sight.

\begin{figure}
\includegraphics[width=0.45 \textwidth]{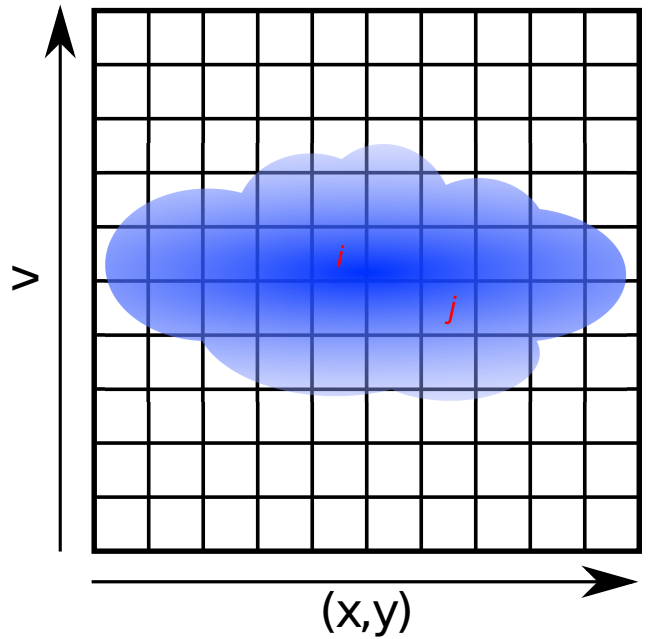}
\caption{Setup of  the problem. We consider a mass
distribution in a 3D PPV data cube. This is represented as the black
grid. $x$--$y$ represents the spatial dimensions and $v$ represents the velocity
dimension. The mass distribution is represented in blue. 
 For each voxel $i$, its  \texttt{G-virial} can be determined by taking its
interactions with all the voxels $j$ (not excluding $i$) into account. See Sect.
\ref{sec:formulation} for  details. }\label{fig:illustration} 
\end{figure}

We start with a mass distribution in a 3D PPV data cube
(Fig.
\ref{fig:illustration}).
We aim to understand how a particle at voxel $i$ is bound by mass from all the
voxels $j$ (not excluding $i$). To achieve this, we split our task into two
subtasks.
First, we need to estimate the gravitational boundedness of voxel $i$ by another
voxel $j$ in the 3D PPV data cube. Second, we need to estimate the gravitational boundedness of a voxel $i$ based on the
information we have about all the other voxels $j$.

\subsection{Boundedness of a voxel pair in a 3D data cube}\label{sec:setup}

We start by defining the gravitational boundedness between a pair of voxels
$(i,j)$.
We are interested in the extent to which a particle $i$ is bound to another voxel
$j$. For our purpose, $i$ can be treated as a massless test particle, and the
boundedness of particle $i$ is determined by the mass of the voxel $j$.

The available measurable quantities in 3D PPV data
cube include the mass $m_j$, the spatial separation $\delta r_{ij}$ and the
velocity separation $\delta v_{ij}$. We expect the boundedness to increase
with $m_j$ and decrease with $\delta r_{ij}$ since gravitational attraction
increases with the mass and decreases with the distance.

The remaining question is how to make use of the information contained in the
velocity direction. We expect that voxels with large velocity differences are
not likely to be bound to each other, for two reasons. First, a
larger velocity separation implies  higher kinetic energy, which consequently
decreases the likelihood of it being gravitationally bound. Second, it has been found
that the molecular gas follows the Larson's relation \citep{1981MNRAS.194..809L}
\footnote{ In \citet{1981MNRAS.194..809L} the index is 0.38. In 
\citet{2011ApJ...740..120R} the index is found to be 0.51.}
where the spatial length and velocity dispersion are related by
\begin{equation}\label{eq:larson}
\delta v \sim \delta r^{1/2}\;.
\end{equation}
 As a result, a larger velocity separation implies a larger
spatial separation, which implies a lower gravitationally boundedness.

Here, we define the gravitational boundedness contributed from voxel $j$ to 
voxel $i$ as
\begin{equation}\label{eq:gvirial:pair}
I_{j\rightarrow i}=\frac{G\; m_j}{\delta r_{ij} \delta v_{ij}^2}\;.
\end{equation}
The boundedness $I_{j\rightarrow i}$ increases with mass
and decreases with spatial and velocity separation.\\

 Apparent similarities
exist between Eq.
\ref{eq:gvirial:pair} and the virial parameter. The virial parameter is defined as 
\citep{1992ApJ...395..140B}\footnote{Here our virial parameter is
$E_{\rm p}/2\; E_{\rm k}$, and in \citet{1992ApJ...395..140B} the virial
parameter is $2\; E_{\rm k}/ E_{\rm p}$. }
\begin{equation}
\alpha_{\rm vir}= \frac{G\; m}{5\; \sigma_{\rm v}^2\; r}\;,
\end{equation} 
where $m$, $r$, and $\sigma_{\rm v}$ are the mass, radius, and velocity dispersion of
the clumps. The virial parameter is a measurement of the gravitational boundedness
of an object. The quantity $I_{j\rightarrow i}$ defined by Eq.
\ref{eq:gvirial:pair} has a dimension of \texttt{G Mass / Radius Velocity$^2$}, which is the same as that of
the virial parameter. Therefore, Eq. \ref{eq:gvirial:pair} can be viewed as a
generalization of the virial parameter to a voxel pair, and a larger
$I_{j\rightarrow i}$ is related to a larger chance for voxel $i$ to be bounded
by voxel $j$.

\subsection{The  \texttt{G-virial} }
We {\it \emph{define}} the  \texttt{G-virial} of a voxel $i$ as the {\it
\emph{sum}} of the gravitational boundedness contributed from all the voxels $j$
(not excluding $i$).
To be more precise,
\begin{align}\label{eq:gvirial}
\alpha_{ \texttt{G-virial}}^{i} &\equiv \sum_{j}  I_{ j\rightarrow i}
\nonumber
\\
&=G \times \sum_{j}  \frac{
m_j((x_i-x_j)^2+(y_i-y_j)^2)^{-\frac{1}{2}}} { (v_i-v_j)^2}  \nonumber
\\
&=G \times \sum_{j}  \frac{
m_j} {((x_i-x_j)^2+(y_i-y_j)^2)^{\frac{1}{2}} (v_i-v_j)^2}  \;,
\end{align}
where  $I_{ j\rightarrow i}$  comes from Eq. \ref{eq:gvirial:pair}, and in the
second step we take into account  that the  \texttt{G-virial} is
measured in the 3D PPV space. The physical meaning of Eq. \ref{eq:gvirial} can
be understood as follows. For one voxel $i$, its  \texttt{G-virial} is determined by summing up its boundedness
with all the other voxels $j$.

In Eq. \ref{eq:gvirial}, if $j$ is close to $i$ in both the
spatial and the velocity direction, it contributes more to the
 \texttt{G-virial}. The contribution is proportional to the mass $m_{j}$, and inversely proportional to $\delta r_{ij}$ and $\delta v_{ij}^2$. 

For a continuous distribution of material, Eq. \ref{eq:gvirial} can be written
as
\begin{multline}\label{eq:gvirial2}
\alpha_{\rm  \texttt{G-virial}}(x,y,v)= \\ 
\int \frac{G \;\widetilde{\rho}(x',y',v')}{((x-x')^2+(y-y')^2)^{1/2}}
\frac{1}{(v-v')^2}\;{\rm d}x' {\rm d}y' {\rm
d}v'\;,
\end{multline}where the integration is carried out over the whole data cube.
 Here $\widetilde{\rho}$ has a dimension of $M\;L^{-2} \; V^{-1}$ where
$M$ is mass, $L$ is size, and $V$ is velocity.

 If  the separation between $i$ and $j$ becomes small, the denominator of Eq.
\ref{eq:gvirial2} approaches zero and the integrand can become large.
If the density distribution is smooth, the term
$$\frac{\widetilde{\rho}(x,y, v)}{((x_i-x_j)^2+(y_i-y_j)^2)^{1/2}}\;,$$
is still finite and does not induce a singularity. However, the term
\begin{equation}\nonumber
\frac{1}{(v-v')^2}\;
\end{equation}
is singular. In our calculations, we change it to
\begin{equation}\label{eq:with:co}
\frac{1}{(v-v')^2+c_{0}^2}\;,
\end{equation}
 which means we suppress the contribution to
the  \texttt{G-virial} when the velocity separation is smaller than $c_0$, and in our case, $c_0$ is comparable to the sound speed. We introduce
 this cutoff in order to make the integrand convergent. Nevertheless, there is a
 physical reason behind this. In a molecular cloud, the velocity dispersion that
 we observe comes from two parts: one  from the thermal motion,
 and the other  from the large-scale streaming motion of the gas. The first
 part is almost scale-independent, and the second part increases with the physical size. We are mainly interested in the balance between gravity and 
 the large-scale streaming motion, and as a result, it is reasonable to
discard the contributions where $\delta_v \lesssim c_0$.
 In our calculations, $c_{0}$ is
 chosen to be $1\;\rm km\;s^{-1}$, which is comparable to the sound speed.
 The effect of changing $c_{0}$ influences the absolute values of the
 \texttt{G-virial}; however, the relative values of the  \texttt{G-virial}
stay unchanged (Appendix \ref{sec:c0}). Therefore, the  \texttt{G-virial} is a
relative measure of the gravitational boundedness. In the case where the
 \texttt{G-virial} map of several observations needs to be compared, a single
value of $c_{0}$ needs to be chosen in advance.

The reasons for the name  \texttt{G-virial} in Eq.
\ref{eq:gvirial} include the following: first, we named it    \texttt{G-virial}
to emphasize its connection with the  commonly used \emph{virial parameter};
second, we added the letter \texttt{G} to emphasize that our virial parameter is
a generalized version of the virial parameter and that it is {\it global}.
Different from the case of \citet{1992ApJ...395..140B} and
\citet{2009Natur.457...63G} where the virial parameter is used to quantify
self-gravity, our  \texttt{G-virial} takes all the gravitational interactions between gas in the data cube into
account. More clarifications concerning the concept of the  \texttt{G-virial} can
be found in Appendix \ref{sec:gvirial:pvirial}.

\section{Separating components with different velocities}\label{sec:separating}
One difficulty that we need to deal with is the line-of-sight confusion. When calculating the  \texttt{G-virial}, different components along the
same line of sight are distinguished through their velocities. If two components
happen to have the same velocity, the  \texttt{G-virial} will be over-estimated,
since physically  unassociated components will be treated as one single component
in the calculations.

We argue that in many
cases, the separation in the velocity axis  provides information so that
we can separate different components based on the velocity difference or
at least alleviate the problem.
We consider two cases; the first  is the line-of-sight confusion in a given
molecular cloud, and the second  is the line-of-sight confusion in our Milky
Way disk.

In a molecular cloud,  the line-of-sight confusion problem can be alleviated
if the molecular gas follows the Larson's relation
\citep{1981MNRAS.194..809L,2011ApJ...740..120R}  because the boundedness is proportional to $I\sim \delta_{\rm r}^{-1} \delta_{\rm v}^{-2}$
(Eq.
\ref{eq:gvirial:pair}) where $\delta_{\rm r}$ and $\delta_{\rm v}$ are the
separations in position and velocity direction, respectively. We consider two points
 along the same line of sight separated by $\delta_{\rm z}$. According to
the Larson's relation, the velocity dispersion (which is a width of the statistical distribution velocity separation
between the two points) is $\delta_{\rm v}\sim \delta_{\rm z}^{1/2}$. Therefore, the
boundedness related to  $\delta_{\rm z}$ by $I\sim \delta_{\rm v}^{-2}
\sim \delta_{\rm z}^{-1}$. In other words, the contribution is not likely to be
large since the velocity is likely to be different in an averaged sense. 

In the case of the Milky Way disk, different spiral arms usually have
different velocities, they will be separated easily in our calculations. We note that in
Eq. \ref{eq:gvirial:pair} the interaction is proportional to $\delta r^{-1}$ and
$\delta v^{-2}$. Therefore, a small separation in velocity will lead to a much
bigger decrease in interaction and finally make the contribution to the total
 \texttt{G-virial} negligible. Since the method uses velocity
difference to deal with line-of-sight confusion, it is more accurate for the cases
where the velocity increases with distance  (e.g., including systematic
expansion and contraction where velocity increases with distance).

\section{Numerical procedure}\label{sec:numerical}
In the case of a 3D PPV data cube, for a voxel $(x,y,v)$, the  \texttt{G-virial}
is defined as (see Eqs. \ref{eq:gvirial2} and \ref{eq:with:co})
\begin{multline}\label{eq:gvirial3}
\alpha_{\rm  \texttt{G-virial}}(x,y,v)= \\ 
\int \frac{G \;\widetilde{\rho}(x',y',v')}{((x-x')^2+(y-y')^2)^{1/2}}
\frac{1}{(v-v')^2+c_{0}^2}\; {\rm d}x' {\rm d}y' {\rm
d}v'\;,
\end{multline}
where the integration is carried out over the whole data cube. With an
observationally constructed distribution of mass in the 3D PPV data cube, we
can easily construct the  \texttt{G-virial} using Eq.
\ref{eq:gvirial3}.

Equation \ref{eq:gvirial3} takes the form of a convolution, and can be conveniently
calculated in Fourier space. In real space, the kernel is 
\begin{equation}
K_{\rm x}=\frac{1}{(x^2+y^2)^{1/2}\;(v^2+c_{0}^2)}\;,
\end{equation}
and in Fourier space it can be shown that\footnote{Here we follow the
non-unitary convention. }
\begin{equation}\label{eq:kernelk}
K_{\rm k}=\frac{2\;\pi}{(k_{\rm x}^2+k_{\rm y}^2)^{1/2}}\; \frac{\pi e^{-c_{0}
|k_{\rm v}|}}{c_{0}}\;.
\end{equation}

Equation \ref{eq:kernelk} enables us to calculate the  \texttt{G-virial} map efficiently.
In our calculations, we first make a 3D Fast Fourier Transform (FFT) to the
observationally constructed mass distribution $\widetilde{\rho}(x,y,v)$ and
obtain $\widetilde{\rho}_{\rm k}(k_{\rm x},k_{\rm y},k_{\rm v})$. 
Then we calculate the  \texttt{G-virial} in the Fourier space
\begin{equation}\label{eq:kernalf}
\alpha_{\rm k}(k_{\rm x},k_{\rm y},k_{\rm v})=\widetilde{\rho}_{\rm k}(k_{\rm
x},k_{\rm y},k_{\rm v})\;\frac{1}{(k_{\rm x}^2+k_{\rm y}^2)^{1/2}}\frac{\pi e^{-c_{0} |k_{\rm v}|}}{c_{0}}\;.
\end{equation}
 In the last step the  \texttt{G-virial} map is obtained by a inverse
FFT of $\alpha_{\rm k}$.

We found that  the absolute values of the
 \texttt{G-virial} are dependent on $c_{\rm 0}$ (see Appendix \ref{sec:c0}); however, the relative values of
the  \texttt{G-virial}  are insensitive to $c_{\rm 0}$. As a result, the
 \texttt{G-virial} is a relative measurement of gravitational boundedness rather
than an absolute one. In order for the  \texttt{G-virial} maps from different
regions to be comparable, a unique value of $c_{\rm 0}$ has to be chosen; $c_{\rm 0}$
also needs to  be larger than the velocity resolution of the observations.

\section{Quantifying the structure of molecular gas
with \texttt{G-virial}}\label{sec:justification}

We present applications of the  \texttt{G-virial} to quantify
gravity and the structure of molecular gas. To begin with, we apply the method
to the models where  both the 3D density and 3D velocity structure are
available. We focus on two aspects:
first, we present the use of  \texttt{G-virial} in quantifying the importance
of gravity in the 3D PPV data cube; second, we quantify
the structures of molecular condensations bases on the
 \texttt{G-virial}.

\subsection{Defining \texttt{G-virial} in the PPP space}
To properly evaluate the accuracy of the \texttt{G-virial} method, we
first define the \texttt{G-virial}$^{\rm model}$ in the PPP space where all the
position and velocity information are available, and then compare it with the
observationally reconstructed values of the \texttt{G-virial} in the PPV space.
Following Equations \ref{eq:gvirial:pair} and \ref{eq:gvirial}, we define the model  \texttt{G-virial} as
\begin{equation}\label{eq:gvirial:model}
\alpha_{ \texttt{G-virial},\; i}^{\rm model}=\sum_{j}I_{j\rightarrow
i}=\sum_{j}\frac{G m_j}{\delta r_{ij} \delta v_{ij}^2}\;,
\end{equation}
where 
\begin{equation}
\delta r_{ij}= ((x_i - x_j)^2+(y_i - y_j)^2+(z_i - z_j)^2 +
\delta_{\rm x}^2)^{1/2}\;,
\end{equation}
and
\begin{equation}\label{eq:3d:dv}
\delta v_{ij}^2= (v_{\rm x}^{i} - v^j_{\rm x})^2+(v_{\rm y}^i - v_{\rm
y}^j)^2+(v_{\rm z}^i - v_{\rm z}^j)^2 + c_{\rm model}^2\;.
\end{equation}
Here the voxel pair $(i,j)$ has positions $((x_i, y_i, z_i), (x_j, y_j, z_j))$
and velocities $((v_{\rm x}^i, v_{\rm  y}^i, v_{\rm z}^i), (v_{\rm x}^j, v_{\rm
y}^j, v_{\rm z}^j))$. The parameter $\delta_{\rm x}$ is the spatial resolution of the
simulation.
Writing Eq.
\ref{eq:gvirial:model} in its integral form, we have
\begin{align}\label{eq:gvirial:3d}
\alpha_{ \texttt{G-virial},\; i}^{\rm model}& = \int \frac{G\; \rho(x', y',
z')}{((x-x')^2 + (y-y')^2 + (z-z')^2 + \delta_{
\rm x}^2)^{1/2}}\times \nonumber \\
& \frac{1}{\splitfrac{ (v_{\rm x}(p) - v_{\rm x}(p'))^2 + (v_{\rm y}(p) -
v_{\rm y}(p'))^2 }{ +(v_{\rm z}(p) - v_{\rm z}(p'))^2 + c_{\rm model }^2)}}
\nonumber
\\
&\times {\rm d}x' \;{\rm d}y'\; {\rm d}z',  
\end{align}
where $p$ stand for $(x,y,z)$ and $p'$ stand for $(x',y',z')$. 
Here we choose $c_{\rm model} =3 \times c_0$ since the denominator of Eq.
\ref{eq:gvirial:3d} is composed of a sum of the velocity difference from the
three dimensions. Since the  \texttt{G-virial} is a relative measure of the
gravitational boundedness (Appendix \ref{sec:c0}), our results are not affected
by this.

\subsection{Quantifying the importance of gravity}\label{sec:gvirial:gvirial}
We study how the importance of gravity traced in the
 \texttt{G-virial} inferred from the simulated observations is related to the
 \texttt{G-virial} theoretically defined in the models. 

We consider two models. The first   ({\it \emph{hydrodynamic simulation model}}) was
taken from a numerical simulation of turbulent gas. The simulation used here was
carried out under periodical boundary conditions, and turbulence is injected through compressive
forcing mode \citep{2008ApJ...688L..79F}, and  no self-gravity is included.
The details of the simulations are described in
\citet{2010A&A...512A..81F,2011ApJ...740..120R}
\footnote{The simulations are available at http://starformat.obspm.fr/starformat/documentation.jsp.}. 
 Since self-gravity is not included in
the simulation,  the line-of-sight confusion effect is probably over-estimated  because, as was  found in \citet{2013ApJ...777..173B}, in the
simulations without gravity structures will overlap more in the PPV space.
In this work, we make use of the density and velocity cubes from the
simulations.

\begin{figure}
\includegraphics[width = 0.45 \textwidth]{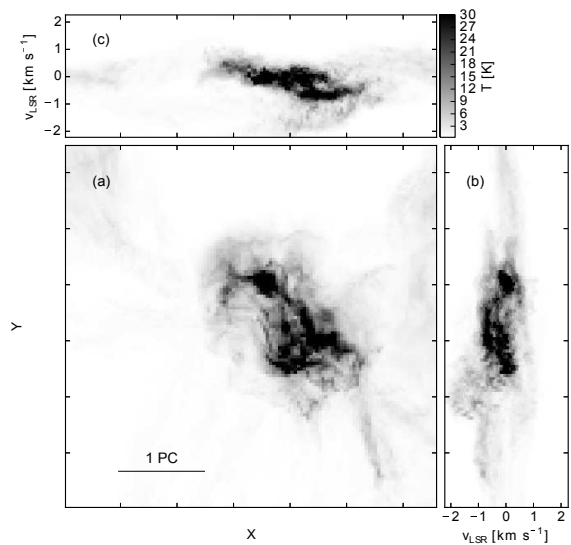}
\caption{ (a) Map of the peak brightness temperature of the simulated
$^{13}$CO(1-0) emission along the velocity axis. (b) Position-velocity map
of peak brightness temperatures of $^{13}$CO(1-0) along the Y-axis (c)
Position-velocity map of peak brightness temperatures of $^{13}$CO(1-0)) along
the X-axis.
}\label{fig:simu:ppv}
\end{figure}

We make a cutout of a size of $4.8{\;\rm pc} \times 4.8 {\;\rm pc} \times 4.8
{\;\rm pc}$ from the snapshot at $t = 5\; T$ where $T$ is the crossing
timescale of the simulation.
The computation of Eq. \ref{eq:gvirial:3d} is computationally expensive ($o(N^2)$).
To save computational time, we choose a portion of the simulation with a
relatively large gas condensation and relatively complicated internal
structures (Fig. \ref{fig:simu:ppv}). 

The second model ({\it \emph{the analytical model}}) is constructed to resemble a
typical molecular condensation \citep[e.g.,][]{2003ApJ...585..850M}. It is defined within
$r_{\rm in}<r<r_{\rm out}$ where $r$ is the radius, $r_{\rm in}=1\;\rm pc$, and
$r_{\rm out}=10 \rm \;pc$. The density structure is \
\begin{equation}
\rho(r)=\big(\frac{r}{r_0}\big)^{\gamma_{\rho}}\times \rho_0\;,
\end{equation}
where $r_0=1\;\rm pc$; $\rho_0=10^3 \times m_{\rm H_2}$, where $m_{\rm H_2}$ is
the mass of the H$_2$ molecule; and $\gamma_{\rho}$ are chosen to vary from
$-1.8$ to $-2.2$.
The velocity structure is parametrized as
\begin{equation}
v(r)=\big( \frac{r}{r_0} \big)^{\gamma_{\rm v}}\times v_0\;,
\end{equation}
where $r_0=1\;\rm pc$,  $\gamma_{\rm v}$ is chosen to vary from $0.35$ to
$0.6$, and  $v_0=1\;\rm km\;s^{-1}$. It can be viewed as an expanding sphere.
 At a given radius, the magnitude
of velocity is distributed uniformly between $0.9\; v(r)$ and $1.1\;v(r)$,
and the direction of the velocity is distributed uniformly in 3D. 
The model is sampled through a total of $3\times 10^5$ particles, and then
turned into a 3D PPV data cube.

We  chose the two models because they represent two typical situations. In the
first case (the hydrodynamic simulation model), the model exhibits a high degree of
physical complexity. In the second case (the analytical model), the density
distribution is centrally condensed, and the model has well-defined
mass-radius and velocity-radius structures. We will study the behavior of the
 \texttt{G-virial} in both cases.

The models are turned into a 3D PPV data cube using a simple radiative transfer
model \citep[Sect.  3.3 of][]{2011ApJ...740..120R}. It is assumed that the
$^{13}$CO(1-0) emission is optically thin and n($^{13}$CO)/n(H$_2$) = 1.7
$\times$ 10$^{-6}$ \citep{1990ApJ...357..477L,1987ApJ...315..621B}. The
excitation temperature is assumed to be 10K. In reality, the $^{13}$CO(1-0)
emission can be optically thick and both the abundance and excitation conditions can vary, and
these can lead to inaccuracies. However, in testing our method,
we are interested in the cases where mass can be reliably traced in the PPV
data cube, and observationally this can be achieved by modeling  several
transitions of CO or using more reliable tracers.

In our calculation of the model  \texttt{G-virial} ($\alpha_{\rm
G-virial}^{\rm model}$), we choose $c_{\rm model} = 1\;\rm km\;s^{-1}$, and in
our calculation of the  \texttt{G-virial} reconstructed from the simulated map ($\alpha_{\rm
G-virial}^{\rm reconstructed}$),
we choose $c_{\rm 0} = 0.3 \;\rm km\;s^{-1}$. We make this choice because in Eq.
\ref{eq:gvirial:3d} all  three velocity components are considered, and in Eq.
\ref{eq:gvirial2} only one velocity component is considered. Since
 \texttt{G-virial} is a relative measure of the gravitational boundedness
(Appendix \ref{sec:c0}), the conclusions are not dependent on this.  
 
 \begin{figure}
    \centering
 \begin{subfigure}[a]{0.5 \textwidth}
 \includegraphics[width = 1 \textwidth]{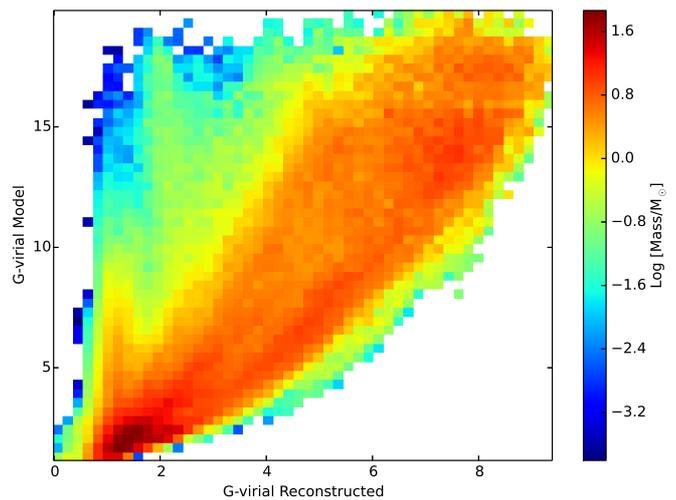}
 \caption{Hydrodynamic simulation.}
 \end{subfigure}
 ~
 \begin{subfigure}[b]{0.5\textwidth}
 \includegraphics[width = 1 \textwidth]{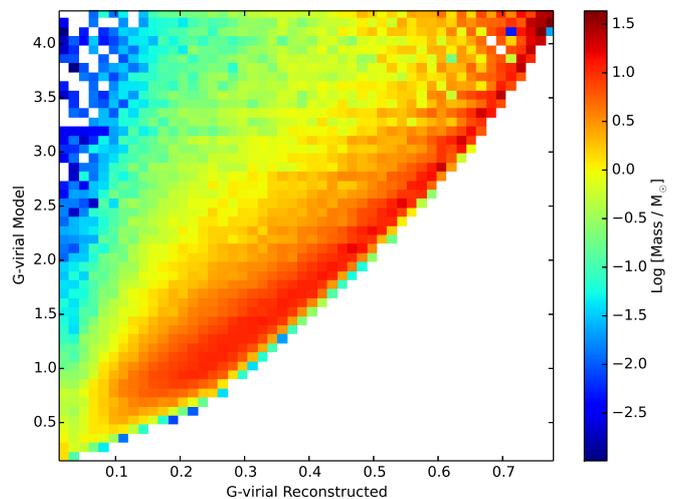}
 \caption{Analytic model.}
 \end{subfigure}

\caption{ A comparison between the  \texttt{G-virial}$_{\rm model}$ calculated
from the model  and the  \texttt{G-virial}$_{\rm reconstructed}$,
which is reconstructed from the simulated observations in the PPV
space. (a) shows the result for the
hydrodynamic simulation and (b) shows
the result for the analytical model with $\gamma_{\rho} = -2.2$ and
$\gamma_{\rm v}=0.5$. The colors stand for
the amount of mass in a given interval.}\label{fig:compare:simulations}
\end{figure}

 In Fig.
\ref{fig:compare:simulations} we compare the  \texttt{G-virial}$_{\rm model}$
calculated from the model with the  \texttt{G-virial}$_{\rm reconstructed}$
reconstructed from the simulated observations in the PPV space. 
The first thing to be noticed is that the absolute values of the
 \texttt{G-virial} differ  because the  \texttt{G-virial} is only a relative measure of the
gravitational boundedness, rather than an absolute one. 

 In
general, a larger  \texttt{G-virial}$_{\rm model}$ is related to a larger
 \texttt{G-virial}$_{\rm reconstructed}$. The reconstructed
 \texttt{G-virial} exhibits a higher uncertainty in the hydrodynamic simulation
model since its flow has complicated structures.

\begin{figure*}
\includegraphics[width = 1\textwidth]{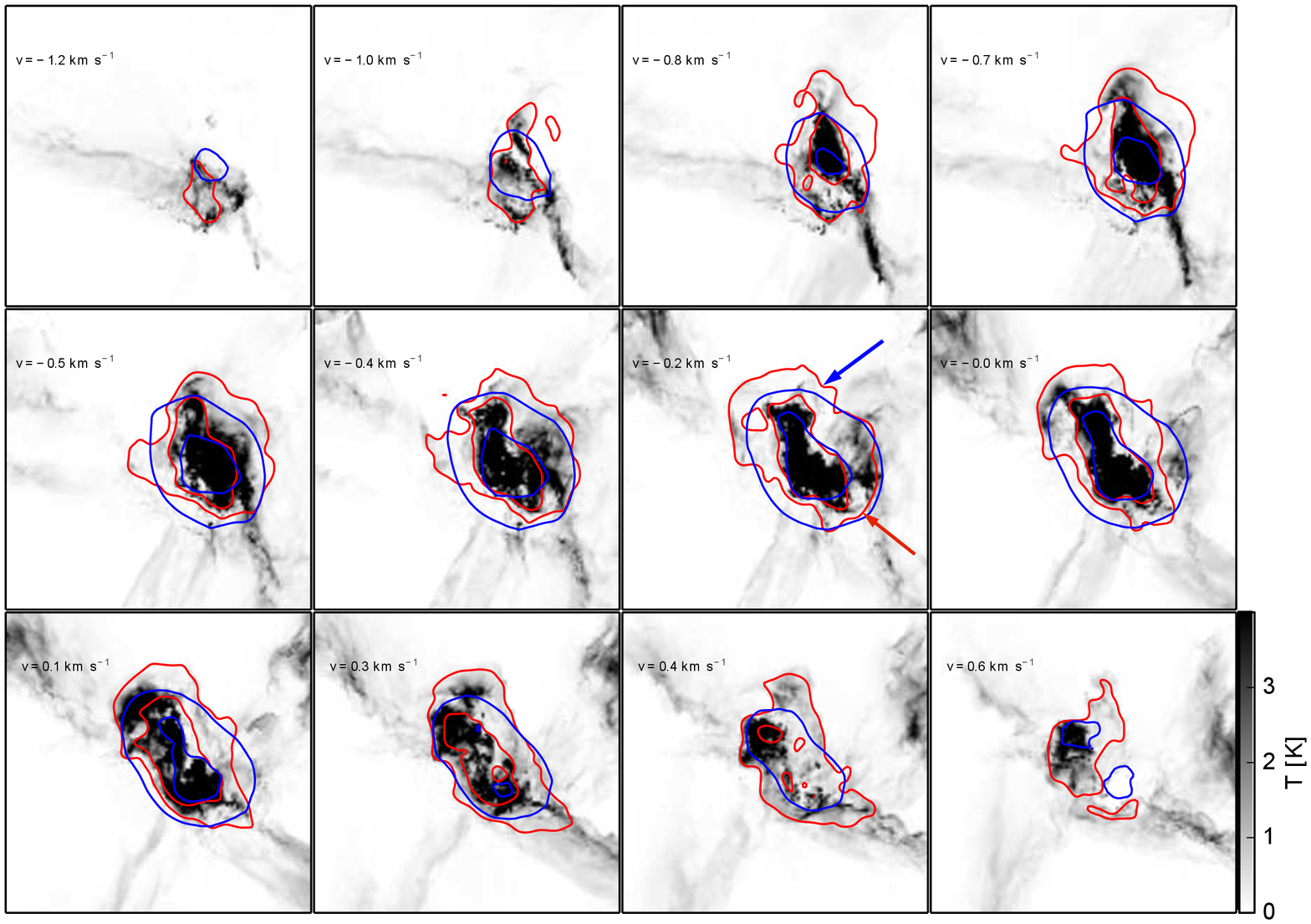}
\caption{\label{fig:gvirial:comp}  A comparison between between 
\texttt{G-virial}$_{\rm model}$ and \texttt{G-virial}$_{\rm reconstructed}$ in the PPV space for the
 hydrodynamical model described in Sect.  \ref{sec:gvirial:gvirial}.
In the model, turbulence is injected through compressible forcing, and no
self-gravity is included \citep{2008ApJ...688L..79F}.
 The
 \texttt{G-virial}$_{\rm model}$ in the PPV space is defined as the mass-weighted
average of the  \texttt{G-virial}$_{\rm model}$ in the PPP
space, which is defined in \ref{eq:gvirial:model}. Here, the
grayscale image is the simulated $^{13}$CO(1-0) emission from the model. The red contours stand for the \texttt{G-virial}$_{\rm model}$, and the contour levels are (6, 5, 10). The blue contours stand for the  \texttt{G-virial}$_{\rm reconstructed}$, and the contour levels are (3, 6). It worth noting that the  \texttt{G-virial}$_{\rm model}$ and the  \texttt{G-virial}$_{\rm reconstructed}$ agrees better at
regions where mass is traced. This can be seen in the $v=-0.2\;\rm km\;s^{-1}$
channel. They agree better at the regions where mass is traced (e.g., the
red arrow), but does not agree well at the region without mass (e.g., the
blue arrow).}
\end{figure*}

To look further  into the relation between  \texttt{G-virial}$_{\rm model}$ and
 \texttt{G-virial}$_{\rm reconstructed}$, we convert the  \texttt{G-virial}$_{\rm model}$ from
the PPP space into the PPV space, and study the connection between the two (Fig.
\ref{fig:gvirial:comp}). At a given velocity interval, the  
 \texttt{G-virial}$_{\rm model}$ in the PPV space is defined as the mass-weighted
average of the  \texttt{G-virial}$_{\rm model}$ in the PPP space, and the average
is carried over the entire line of sight.
There are some significant differences between the iso- \texttt{G-virial}$_{\rm
model}$ contours  (isosurfaces where the  \texttt{G-virial} takes constant
values) and iso- \texttt{G-virial}$_{\rm reconstructed}$ contours:
the iso- \texttt{G-virial}$_{\rm model}$ exhibit more structures and
iso- \texttt{G-virial}$_{\rm reconstructed}$ is  smoother  because the  \texttt{G-virial}$_{\rm model}$ is unaffected by the
line-of-sight confusion. Thus, the  \texttt{G-virial} map constructed in the PPV
space is better suited to studying the effect of gravity at large scale, but is not
suitable for the study of the importance of gravity at small scale (e.g.,
individual cores).
Although the iso- \texttt{G-virial}$_{\rm model}$ and the
iso- \texttt{G-virial}$_{\rm reconstructed}$ does not agree completely, they
agree better at positions where emission is enhanced and mass is traced, and
tends to differ more where there is little emission. Since we are interested in
the gravitational boundedness of the gas, this disagreement at regions where
little mass is present is not as important as it might appear.

Since the
hydrodynamic simulation model exhibit complicated structures,
the line-of-sight confusion has a larger impact on the results than the
case of the analytical model. However, in the  gravity-free hydrodynamic
simulation model the confusion probably has been over-estimated. According to
\citet{2013ApJ...777..173B}, structures tend to overlap more without
self-gravity.

\subsection{Quantifying the internal structures}\label{sec:model:internal}

 For a given region, its  \texttt{G-virial} map is centrally concentrated,
and if we divide the region based on different isosurfaces of
 \texttt{G-virial} values, we get a set of regions that nest inside one another. 
We propose to quantify the structures of the gas condensations in the PPV space
with the mass-radius relation and velocity dispersion-radius relation derived with the help of the
isosurfaces of the  \texttt{G-virial}. 

Inside a closed region where
the  \texttt{G-virial} of the voxels is larger than a given threshold, we can
evaluate parameters such as mass, radius, and velocity dispersion. 
Using a map of mass distribution $m(x,y,v)$ and a map of the
 \texttt{G-virial} $\alpha_{\rm  \texttt{G-virial}}(x,y,v)$, the corresponding
parameters are evaluated by taking all the voxels inside a  given contour of
the $\alpha_{\rm  \texttt{G-virial}}(x,y,v)$ map into account. 
The mass of a region is defined as
\begin{equation}\label{eq:def:mass}
m=\int_R \widetilde{\rho} (x,y,v) \;{\rm d} x  {\rm d} y  {\rm d} v\;,
\end{equation}
where $R$ denotes a coherent region that satisfies $\alpha_{\rm
 \texttt{G-virial}}(x,y,v)>\alpha_{\rm min}$
and the central velocity of a region is defined as 
\begin{equation}
v_0 = \frac{\int_R \rho (x,y,v)\; v\; {\rm d} x  {\rm d} y  {\rm d} v}{\int_R
\rho (x,y,v) \; {\rm d} x  {\rm d} y  {\rm d} v}\;.
\end{equation} 
The velocity dispersion is defined as 
\begin{equation}\label{eq:sigmav}
\sigma_{\rm v}= \Big(\frac{\int_R \rho (x,y,v) \;(v-v_0)^2\; {\rm d} x  {\rm d}
y {\rm d} v}{\int_R \rho(x,y,v)\; {\rm d} x  {\rm d} y  {\rm d} v}
\Big)^{1/2}\;.
\end{equation}
The radius of the region is defined by diagonalizing the tensor of second
moments of the position coordinates weighted by the intensity
\citep{2009Natur.457...63G}, and is defined as 
\begin{equation}\label{eq:def:radius}
R=1.91 \times (\sigma_{\rm min}\sigma_{\rm max})^{1/2}\;,
\end{equation}
where  $\sigma_{\rm min}$and  $\sigma_{\rm max}$ are the dispersions along the
major and minor axes.
  
This set of definitions enables us to quantify the structure of molecular
condensations observed in the PPV space in terms of mass-radius and velocity
dispersion-radius relations. If the molecular gas has a relatively diffuse
morphology, we expect to see a steeper dependence of the enclosed mass $M$ with
the radius $r$; if the molecular gas is centrally condensed, we expect to see a
shallower $M$-$r$ dependence. Similar arguments can be applied to the dependence
of velocity dispersion on radius $\sigma_{\rm v}$-$r$.

\begin{figure*}
\includegraphics[width=0.49 \textwidth]{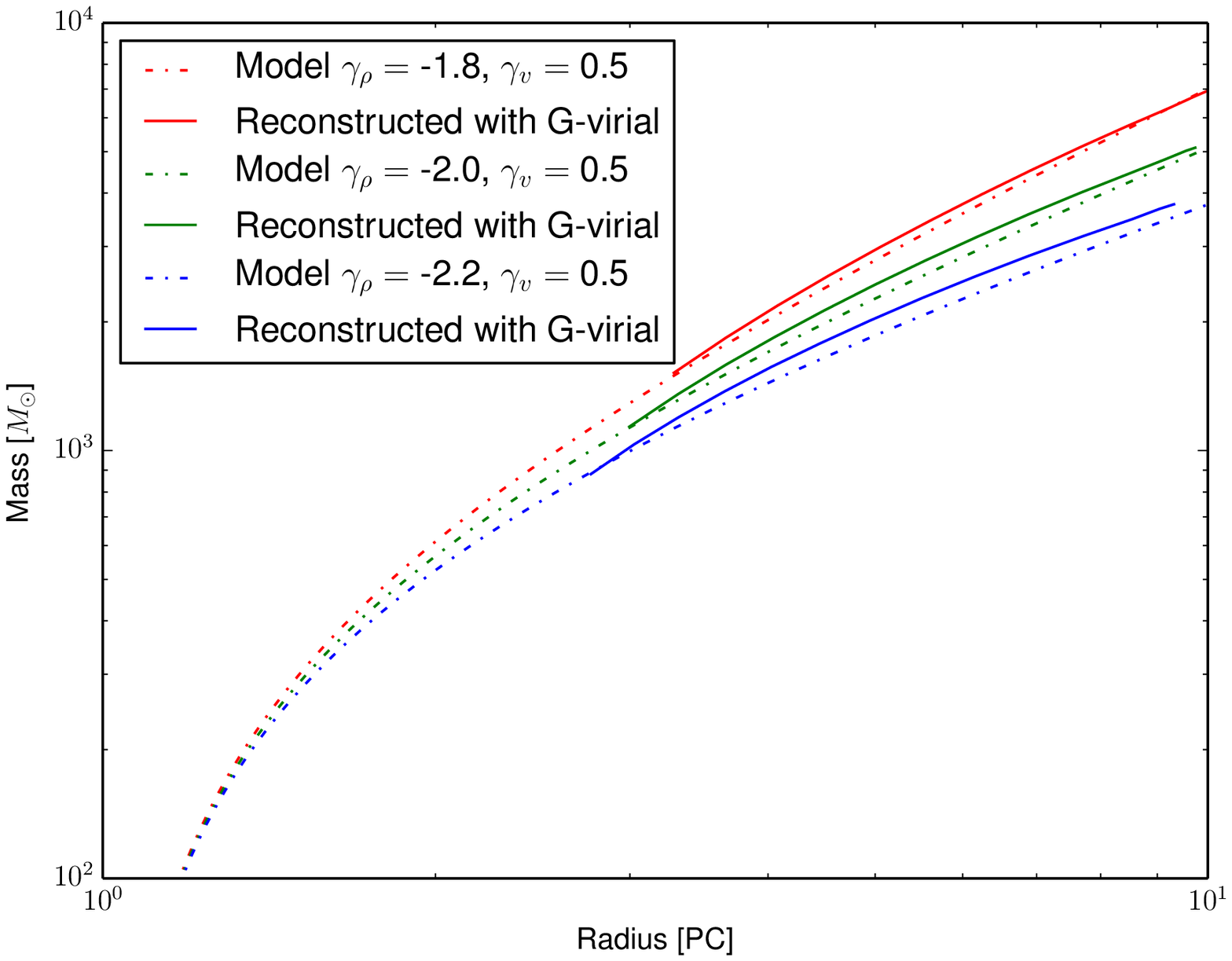}
\includegraphics[width=0.49 \textwidth]{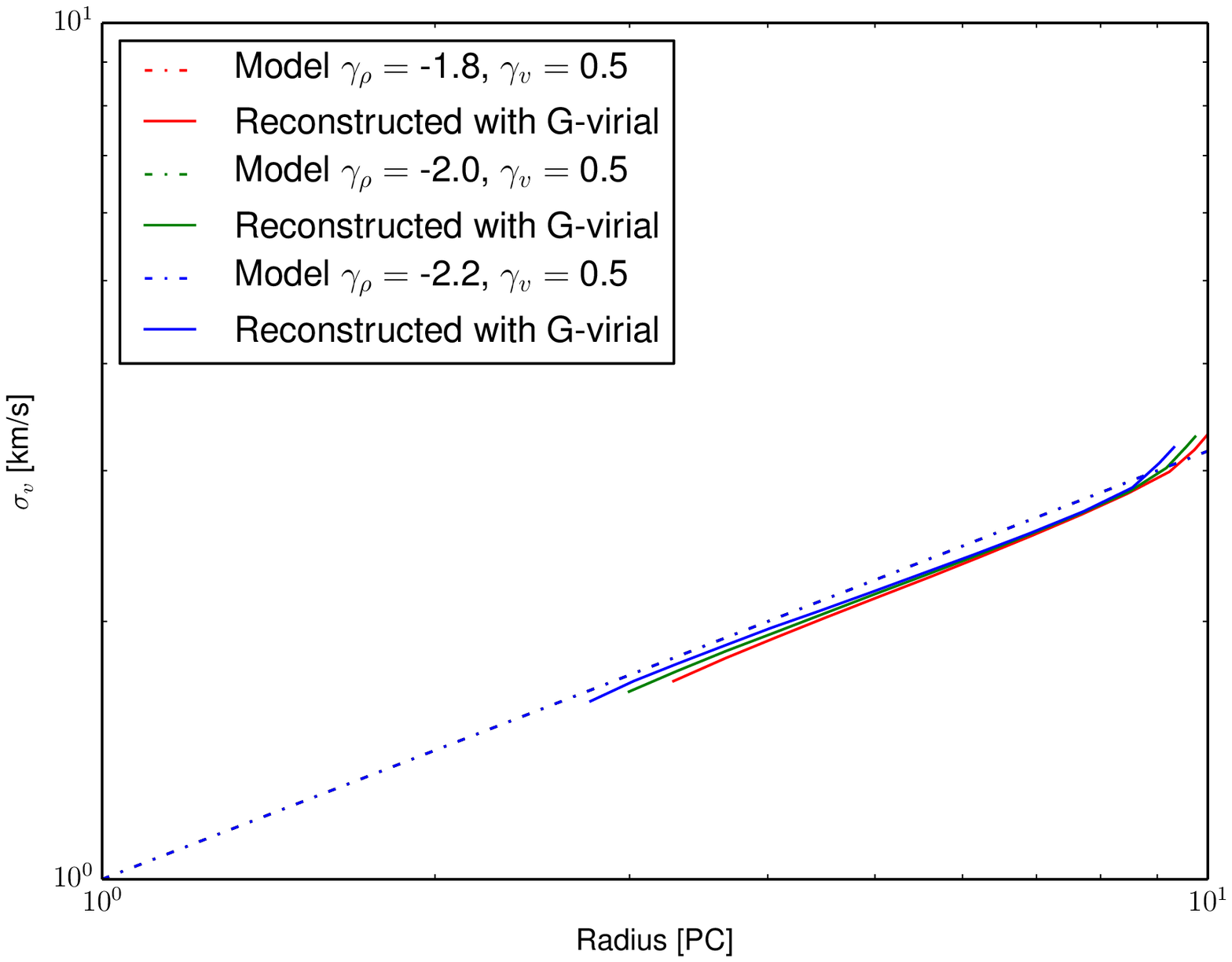}
\includegraphics[width=0.49 \textwidth]{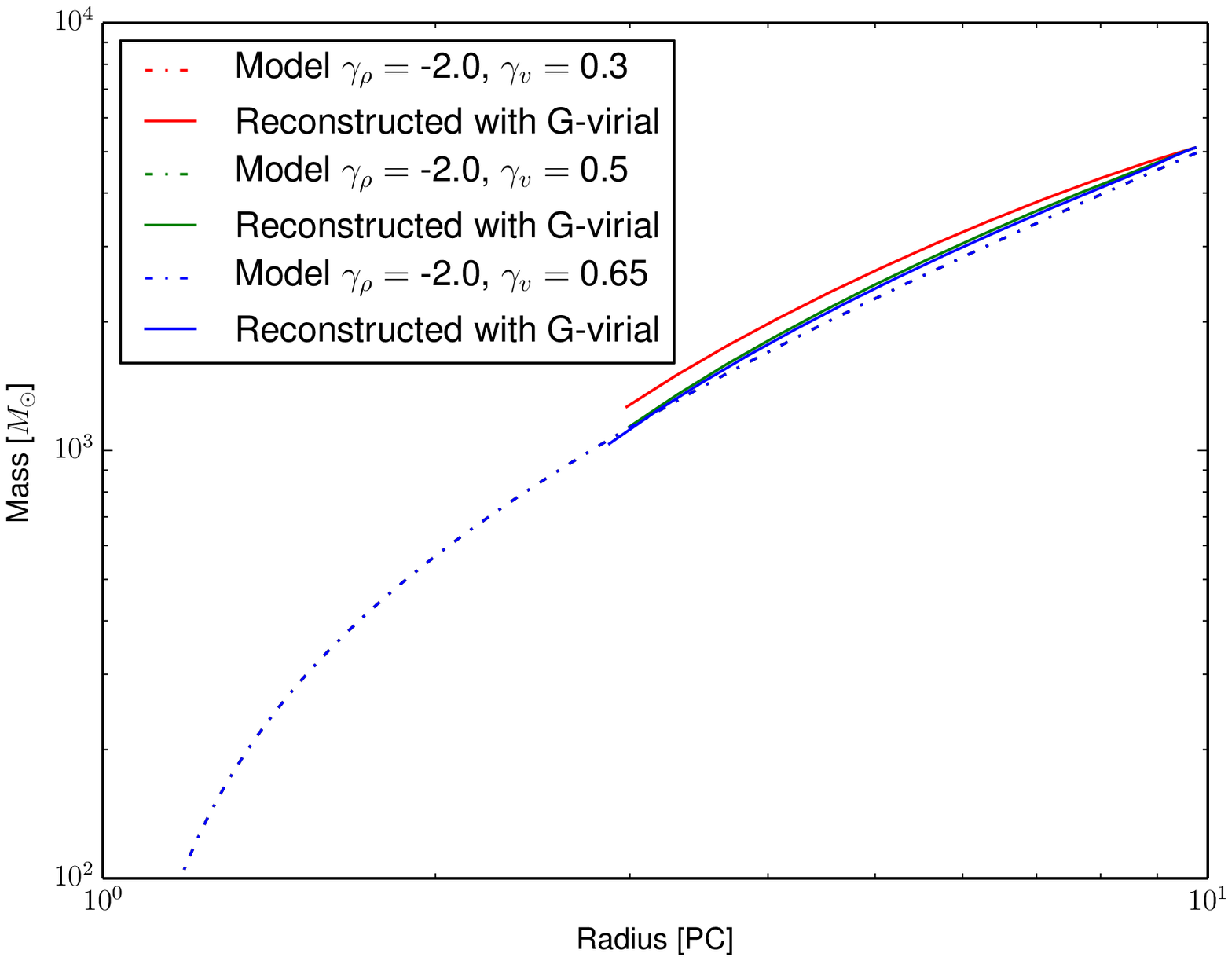}
\includegraphics[width=0.49 \textwidth]{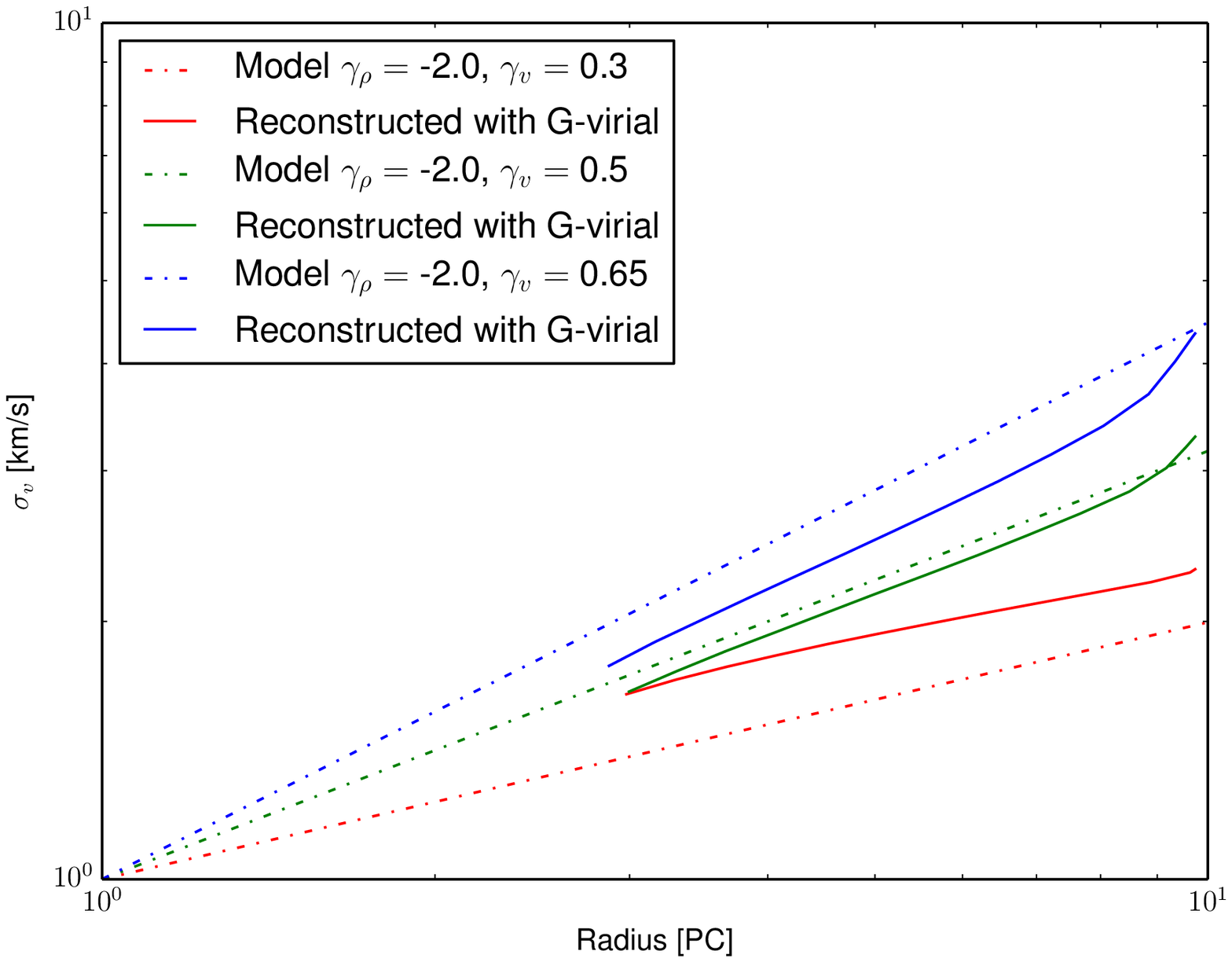}
\caption{ Comparison between the structure of the model and the structure
reconstructed through the  \texttt{G-virial} for different sets of model
parameters. The upper panels show the results where the analytical model is
constructed with different $\gamma_{\rm \rho}$, and the lower panels show the
results where the analytical model is constructed with different $\gamma_{\rm
v}$. 
 The left panels are the mass-radius relations and the right panels are the
velocity dispersion-radius relations. Here the radius is scaled up by a factor
of 1.4 and the velocity dispersion is scaled up by a factor
of 1.45 for an easy comparison. }\label{fig:compare:analy}
\end{figure*}

If the molecular gas is centrally condensed, the $M$-$r$ and $\sigma_{\rm
v}$-$r$ relations of the gas condensations are well defined in the
PPP (position-position-position) space.
It is expected that we can reconstruct these relations observationally with the
help of the  \texttt{G-virial}. If a molecular gas condensation has a diffuse
and irregular morphology, these  \texttt{G-virial}-derived relations can still
represent the morphology of the gas condensations. The difference between
different regions can be represented in the $M$-$r$ and $\sigma_{\rm
v}$-$r$ planes.

To demonstrate the diagnostic power of the $M$-$r$ and $\sigma_{\rm v}$-$r$
relations, we generated a set of models with different density and velocity
structures, and compare them with the $M$-$r$ and $\sigma_{\rm v}$-$r$ relations
reconstructed with the  \texttt{G-virial }(Fig.
\ref{fig:compare:analy}). 
The differences in structure among different regions can be accurately reconstructed with the help of
the  \texttt{G-virial}. Furthermore, the slope of the $M$-$r$ and $\sigma_{\rm v}$-$r$
relations are accurately reconstructed with the  \texttt{G-virial}
\footnote{To make the  \texttt{G-virial}-derived relations overlap with the model
values, we scaled the estimated radius by a factor of 1.4 and the estimated velocity
dispersion by a factor of 1.45. This adjustment is made in order to make it
easier to compare different curves. In practice, the radii of the estimated
regions are based on the tensor of second
moments of the position coordinates, and the radii do not necessarily coincide
with the radii defined in the model. A similar argument also applies to the estimation of the velocity dispersion. }.

\begin{figure}
\includegraphics[width=0.49 \textwidth]{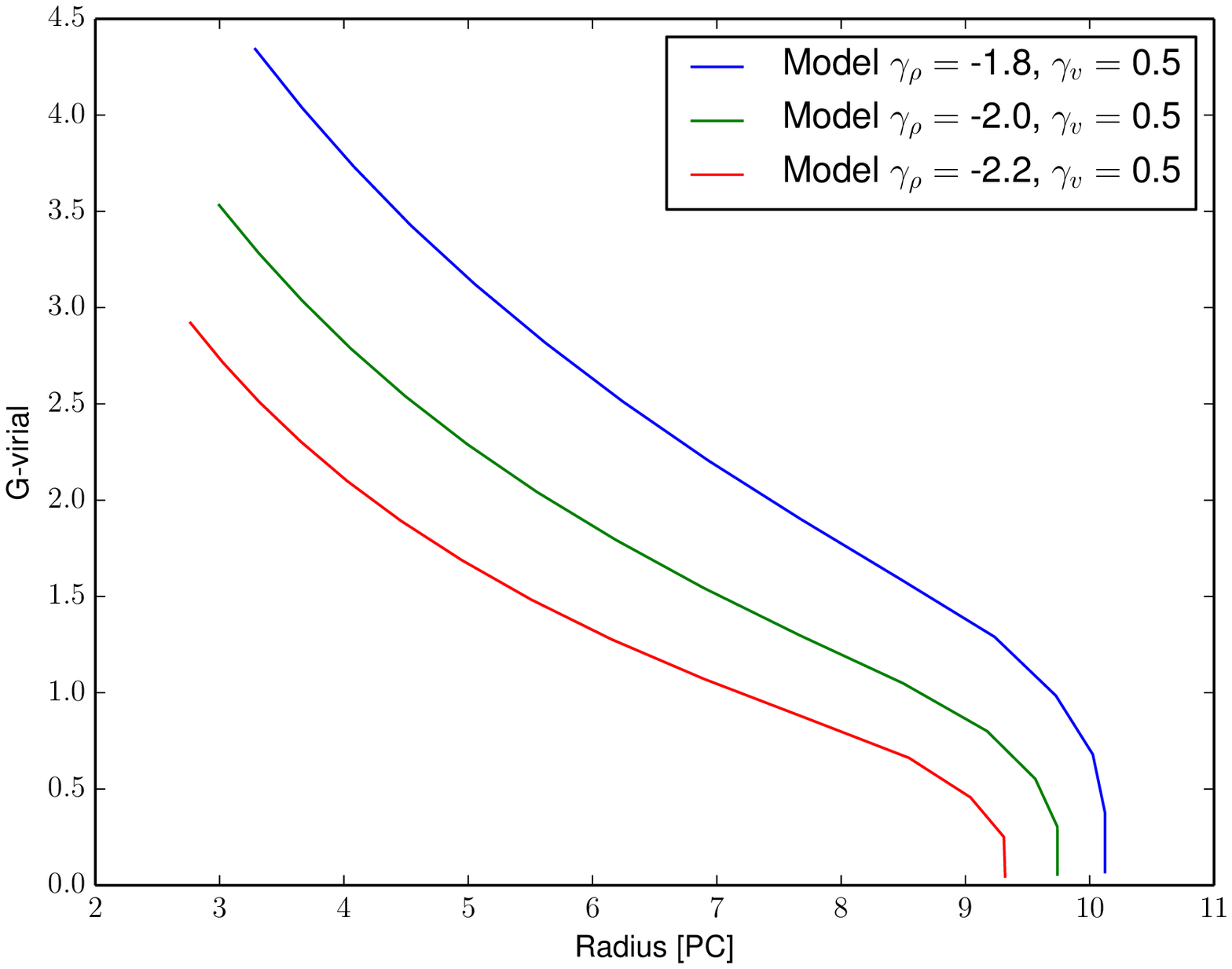}
\includegraphics[width=0.49 \textwidth]{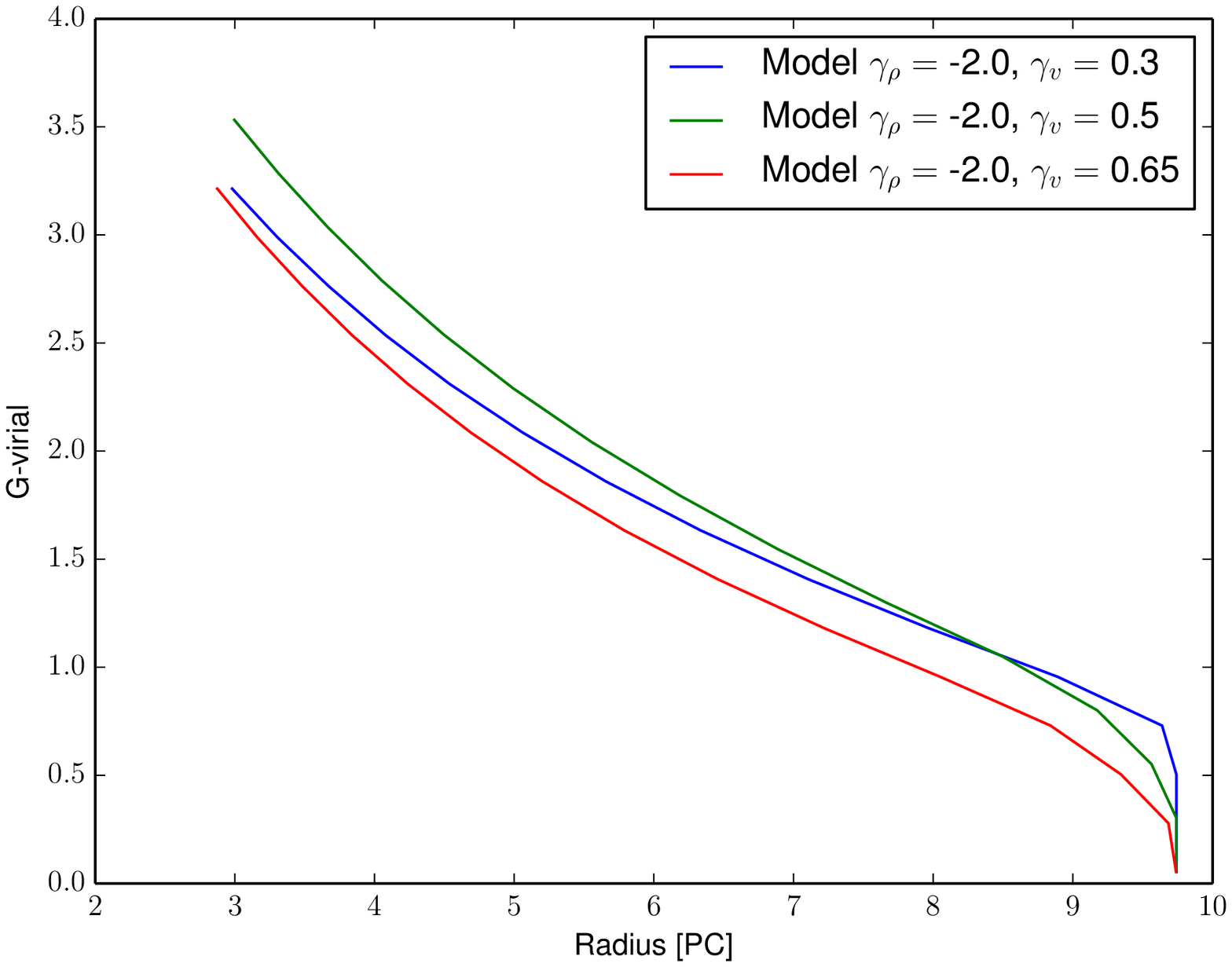}

\caption{ \texttt{G-virial} as a function of radii for models with different
$\gamma_{\rho}$ and $\gamma_{v}$. }
\label{fig:compare:gvirial}
\end{figure}

We also plot the dependence of $ \texttt{G-virial}$ as a function of radius for
various models (Fig. \ref{fig:compare:gvirial}). For all the models, we found an increase of
 \texttt{G-virial} towards the centers of the regions. Larger  \texttt{G-virial}
are found for the models with more condensed structure and lower velocity
dispersion.

\subsection{Short summary}
In this section we demonstrated the usage of  \texttt{G-virial} to quantify the
structure of molecular gas condensations. We summarize our findings and
remind the reader of the caveats. 

We found that the  \texttt{G-virial} can be used to quantify the importance
of gravity and to quantify the structures of the gas condensations. 
\begin{enumerate}
  \item The  \texttt{G-virial} measured from the PPV data cube is
  positively related to the  \texttt{G-virial} found in the model (PPP
  space), and a larger  \texttt{G-virial} is related to a larger probability of
  being gravitationally bound by the ambient gas.
  \item The  \texttt{G-virial} map can be used to derive mass-radius and
  velocity dispersion-radius relations.
  The  \texttt{G-virial}-derived relations are good reconstructions of the
  structures of the models when the models are symmetric,
  and can also provide good representations of the structures if the models
  are irregular.
\end{enumerate}

Several caveats need to be noted. First, the  \texttt{G-virial} is a relative
measure of the gravitational boundedness. A larger  \texttt{G-virial}
means a larger change of being gravitationally bound. The absolute values of
the  \texttt{G-virial} cannot be directly used to tell if a region is
gravitationally bound or not. The absolute values of the
 \texttt{G-virial} are dependent on $c_0$. Therefore, in order for the values of
 \texttt{G-virial} from several maps to be comparable, a unique value of $c_0$
needs to be chosen in advance. This value should be comparable to the sound speed of
the medium, and in this work we have chosen it to be $c_0 = 1 \rm km \; s^{-1}$.
We also note that the  \texttt{G-virial} is reconstructed from the PPV space where the projection effect tends to smear out
structures; it is more suitable for the study of the gravitational boundedness
of the structures larger than the individual gas condensations. 

Second, the  \texttt{G-virial} method can be used to derive the mass-radius and
velocity dispersion-radius relations for molecular condensations. However, the
results need to be properly interpreted. If the structures of the gas
condensations are regular and are close to being spherically symmetric, the
G-virial-derived mass-radius and velocity dispersion-radius relations should be
good reconstructions of the real 3D structure. If the structure of
the gas condensations are irregular, the  \texttt{G-virial} can be used to derive
these relations.  However, in this case, the \texttt{G-virial}-derived
relations should be viewed as representations of the
compactness of the structures in the PPV space, and are not necessarily related
the compactness of the structures in the PPP space.

Finally, we would like to emphasize that the  \texttt{G-virial} methods works in
the PPV space, and like other methods, its accuracy is limited by
line-of-sight contamination. The method relies on using the velocity difference
to alleviate the line-of-sight confusion. This requires
different gas components to have different velocities, which is satisfied in
many cases.
For a single object, this requires the velocity separation to increase with radius. Whereas this generally holds for
individual molecular clouds and clumps, it does not hold for some cores or disks
where the velocity decreases with radius (e.g., collapsing cores).

\section{Applications to molecular clouds}\label{sec:appl}
In this section we present applications of our method to several molecular
clouds. Our method requires a map of the distribution of molecular gas in
the 3D PPV space. Observationally, this can be
conveniently achieved using rotational transitions of the CO molecule.

In the simplest case, this can be achieved by observing the $^{13}$CO(1-0)
transition alone. Assuming an excitation temperature of 10 K, a $^{12}$CO
abundance of X($^{12}$CO/H$_2$)=$8\times 10^{-5}$, and a $^{12}$CO to $^{13}$CO ratio of
R($^{12}$CO/$^{13}$CO)=45, the column density can be estimated as
\citep{2001ApJ...551..747S,2010ApJ...723..492R}
\begin{equation}
N(H_2)=4.92\times 10^{20}\times \frac{T_{\rm mb}}{\rm K}\times\frac{\delta_{\rm
v}}{\rm km\;s^{-1}} \;,
\end{equation}
where $T_{\rm mb}$ is measured in K and $\delta_{\rm v}$ is the velocity resolution of
the data cube measured in $\rm km\;s^{-1}$.
Changing the excitation temperature to 20 or 30 K decreases the derived mass by
40\% and 92\%, respectively \citep{2001ApJ...551..747S}. For an observed
$^{13}$CO(1-0) data cube $T_{\rm mb}(x,y,v)$, the mass distribution $m(x,y,v)$
can be obtained as
\begin{multline}
\frac{m(x,y,v)}{ g}=4.92\times 10^{20} \;\times \frac{T_{\rm mb}(x,y,v)}{\rm K}
\frac{\delta_{\rm v}}{\rm km \;s^{-1}}  \\
 \times \frac{m_{\rm H_2}}{\rm g}\times \Big(\frac{\delta_{\rm x}}{\rm
 cm}\Big)^2\;,
\end{multline}
where $\delta_{\rm x}$ is the voxel size of the spatial dimension in $\rm cm$
and $\delta_{ \rm v}$ is the channel width in $\rm km\;s^{-1}$. The values of $m$ and
$\widetilde{\rho}$ in Eq. \ref{eq:gvirial2} are related by  $m =
\widetilde{\rho}\; {\rm d}x \; {\rm d}y\; {\rm d} v$.

We apply our  \texttt{G-virial} method to the publicly available data from the
COMPLETE survey \citep{2006AJ....131.2921R}.
The observations have a spatial resolution of $46\arcsec$ and a velocity resolution of
$0.067\;\rm km\;s^{-1}$, the mean RMS per channel is $\sim 0.33 \;\rm K$ in
terms of $T_{\rm A}^*$, and the beam efficiency is $\sim 0.5$. 

 With a distance of 250$\pm$50 pc and a total
mass of $10^{4}M_{\odot}$ \citep{2006ApJ...638..293E}, the Perseus molecular cloud is among the
best-studied molecular clouds in the Milky Way. It is composed of several
distinct regions: B1, B3, NGC1333, and IC348. The Ophiuchus molecular
cloud has a distance of 125pc and a mass of $7\times10^3\;M_{\odot}$
\citep{1989A&A...216...44D} and contains clouds L1688 and L1689.
In particular, L1688 and NGC1333 are cluster-bearing regions. IC348 also
hosts a star cluster; however, it is generally considered older than the other
regions in Perseus \citep{2009ApJS..184...18G}.

In this section, we provide a study of the structure of the Perseus and
Ophiuchus molecular clouds with our  \texttt{G-virial} method. We present maps of the
 \texttt{G-virial}, together with an analysis of the structure of
the regions in the clouds based on the  \texttt{G-virial} method.

\subsection{Maps of the G-virial}
\begin{figure*}
\includegraphics[width=0.95\textwidth]{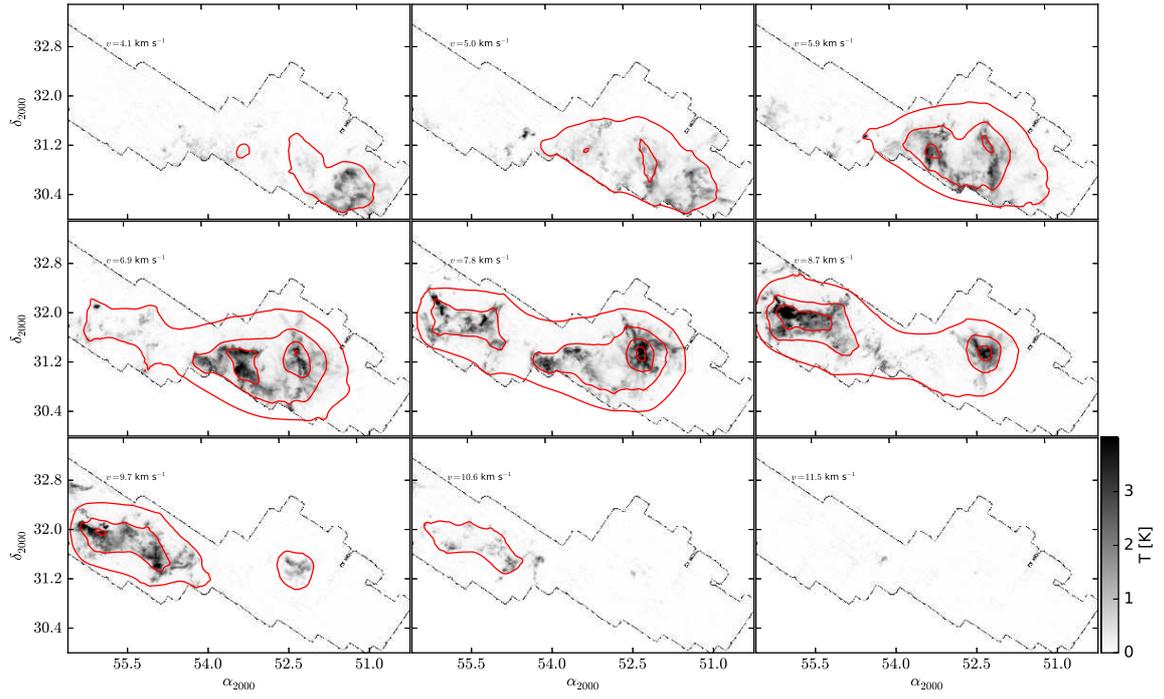}

\includegraphics[width=0.95\textwidth]{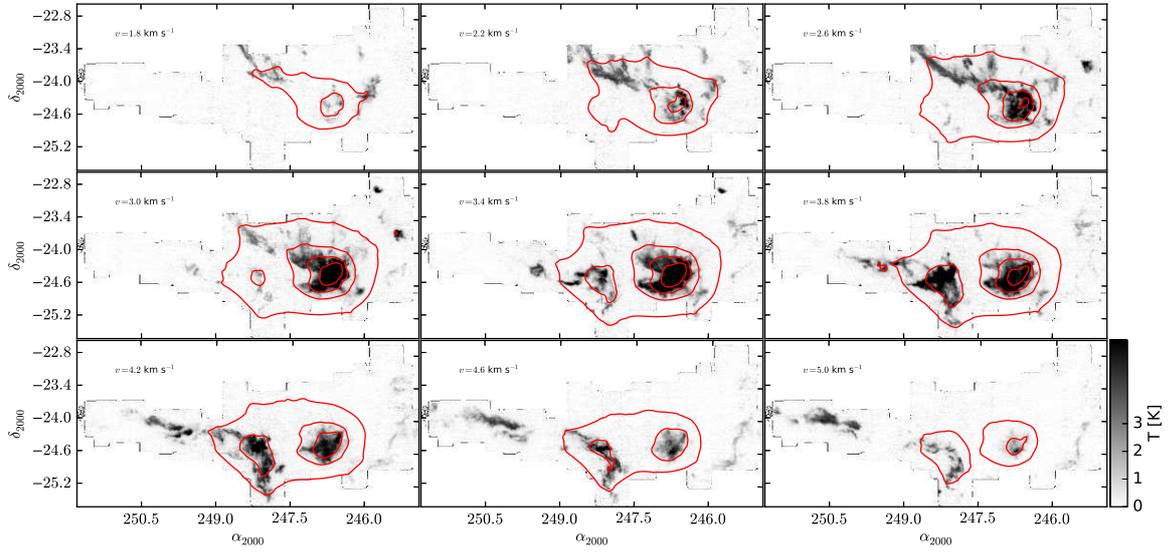}

\caption{\label{fig:channel} 
Channel maps of the
$^{13}$CO(1-0) emission and the corresponding  \texttt{G-virial}. The upper
panel shows the results from the Perseus molecular cloud and the lower panel
shows the results from the Ophiuchus molecular cloud. Grayscale images stand for
$^{13}$CO(1-0) emission and the red contours represents the
 \texttt{G-virial}. Contour levels start from 1.2 in steps of 0.8. }
\end{figure*}

In Fig. \ref{fig:channel} we channel maps of the  \texttt{G-virial} in the 3D
PPV space.  Three-dimensional renderings of the maps are presented in
Appendix \ref{sec:rendering}. For both clouds, the $^{13}$CO(1-0) maps contain
complicated and filamentary structure, whereas the  \texttt{G-virial} maps are smooth and contain fewer structures. Interestingly,
higher values of  \texttt{G-virial} are reached only at cluster-bearing
regions such as NGC1333, IC348, and L1688, and this highlights the importance of
gravity in such regions. Different regions are indicated in Fig.
\ref{fig:regions}. 

\subsection{Identification of regions}
The  \texttt{G-virial} offers a new way to divide the molecular cloud into
regions.
Previously, this has been done either by visual inspection, or with
contour-based algorithms such as \texttt{clumpfind} \citep{1994ApJ...428..693W}. It is worth
mentioning that other region-finding algorithms are available, such as
\texttt{dendrogram} \citep{2008ApJ...679.1338R} and \texttt{dochamp} \citep{2012MNRAS.421.3242W}.

Many of the algorithms (e.g., \texttt{clumpfind} and \texttt{dendrogram}) are
contour-based, and they tend to assign voxels to regions based on iso-column-density contours
in the 2D case or on iso-intensity contours in the case of a 3D
PPV data cube. Molecular clouds are characterized by a
set of complicated hierarchical structures, and a naive application of the
\texttt{clumpfind} tends to produce hundreds of clumps for one single
molecular cloud.
It has also been pointed out recently that contour-based methods suffer from
superposition and confusion when the volume filling factor of emitting material
is large \citep{2013ApJ...777..173B}.

\begin{figure}
\includegraphics[width=0.45\textwidth]{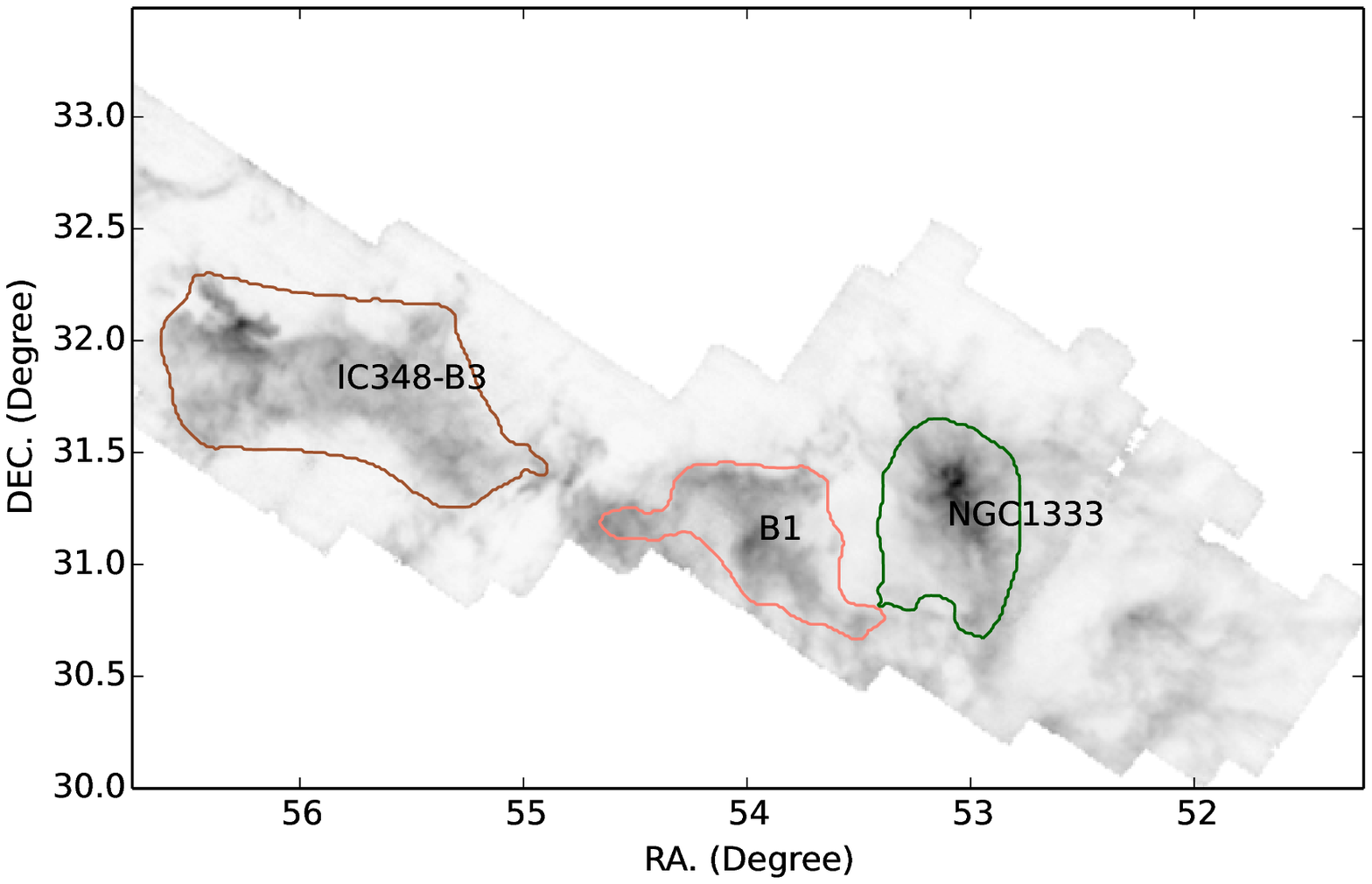}

\includegraphics[width=0.45\textwidth]{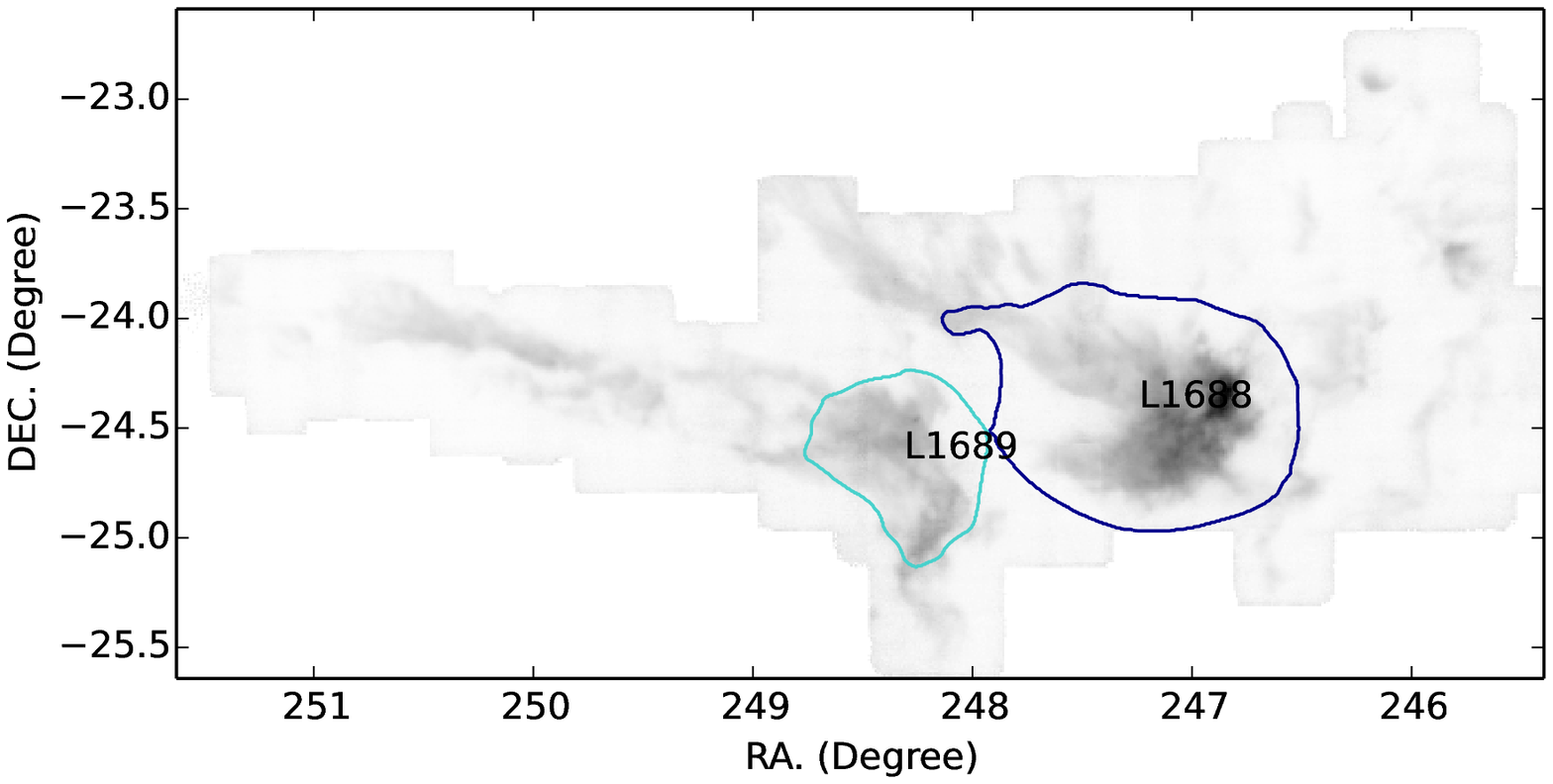}

\caption{\label{fig:regions} Regions identified by applying the
\texttt{dendrogram} algorithm to our  \texttt{G-virial} map. 
The results from the Perseus molecular cloud (upper panel) and Ophiuchus
molecular cloud (lower panel) are presented.  The grayscale
images correspond to the velocity-integrated $^{13}$CO(1-0) emission and the
contours correspond to the projected boundaries of the identified regions. These
regions correspond to the ``leaves'' of the \texttt{dendrogram}. The conventional names of the regions are
labeled.
}
\end{figure}

In our analysis we are interested in finding gravitationally coherent regions.
 This can be achieved by applying the \texttt{dendrogram} to the  \texttt{G-virial}
map, which is a measurement of the gravitational boundedness.
Figure \ref{fig:regions} shows the regions identified by applying the
\texttt{dendrogram} algorithm to our  \texttt{G-virial} map. 
 The identified regions correspond to the ``leaves'' of the
\texttt{dendrogram}.  In the
\texttt{dendrogram} calculations, the minimum difference between different
contours is set to 0.4, and the regions correspond to the leaves of the
\texttt{dendrogram}. Each identified region consists of a continuous list of voxels in the 3D
PPV space. In Fig.
\ref{fig:regions} the projected boundaries of the regions are plotted
\footnote{All the regions were originally defined in 3D. }.

Compared to \texttt{clumpfind}, a combination of the  \texttt{G-virial} method
and the \texttt{Dendrogram} method tends to identify regions that are coherent
while the \texttt{clumpfind} tends to break up those regions.
The reason is that our \texttt{ \texttt{G-virial}} maps are much smoother than
the original map \footnote{This is similar to the case of
\citet{2009MNRAS.396..830S} where they identified structures form a simulation
in the 3D position-position-position space based on gravitational potential.}.
Since the  \texttt{G-virial} maps are much  smoother, the results obtained with
the  \texttt{G-virial} maps are much less sensitive to the technical parameters
 (e.g., the minimum separation of the contours) used in the computations
compared to the case with the intensity maps.
Therefore, the  \texttt{G-virial} offers a more robust definition of regions.
With  \texttt{G-virial} we can study the structure of molecular gas in terms of these {\it \emph{gravitationally coherent}} regions. In the 2D case, our method is similar to the \texttt{gridcore} code which is based on 2D projected gravitational potential \citep{2011ApJ...729..120G}.

\subsection{Internal structure of the regions}
In this section we study the internal structure of the molecular gas in the
individual regions. Previously, the structure of molecular gas was
studied in terms of the clumps, and quantified using parameters that are
evaluated for the whole clumps. While these methods do provide constraints on
the role of gravity inside the regions, the role of gravity on scales
larger than the individual gas condensations is neglected. The spatial structure of
molecular gas larger than the clump scale can be quantified using the
\texttt{dendrogram} algorithm. However, the \texttt{dendrogram} method is contour-based, and only the connection between
the adjacent regions in the tree diagram are preserved.
This makes it difficult to use \texttt{dendrogram} to quantify the structure of
molecular gas on the large scale. 

 Here we provide an analysis of the structure of the individual
regions with the  \texttt{G-virial} method. As we have shown in section
\ref{sec:model:internal}, the intrinsic mass-radius and velocity-radius relations
in the model can be reconstructed with our
 \texttt{G-virial} method   to a good accuracy if the velocity dispersion
 increases with the size.

\begin{figure}
\includegraphics[width=0.45\textwidth]{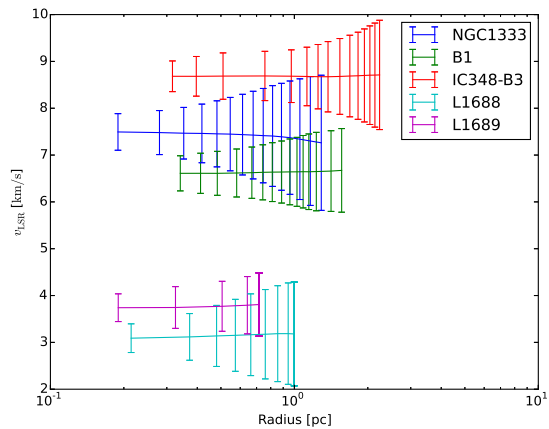}
\caption{\label{fig:vr}Dependence of velocity and velocity dispersion as a
function of radius. Here the solid lines represent the velocities of the
regions, and the bars represent the velocity dispersions of the regions at
given radii.
The vertical extent of the errorbars are the 3D velocity dispersion
$\sigma_{\rm v}^{\rm 3D}=\sqrt{3}\;\sigma_{\rm v}$. }
\end{figure}

In Fig. \ref{fig:vr} we plot the dependence of velocity $v_0$ and
velocity dispersion $\sigma_{\rm v}$ on the radii of the regions. 
The inner parts of the regions are clearly at the center of the outer part of the 
regions in velocity space.
This means the central part of the regions remains quiescent with respect to the
outer parts, and this is consistent with the findings by
\citet{2010ApJ...723..457K,2004ApJ...614..194W,2007A&A...472..519A}.

\begin{figure}
\includegraphics[width=0.45\textwidth]{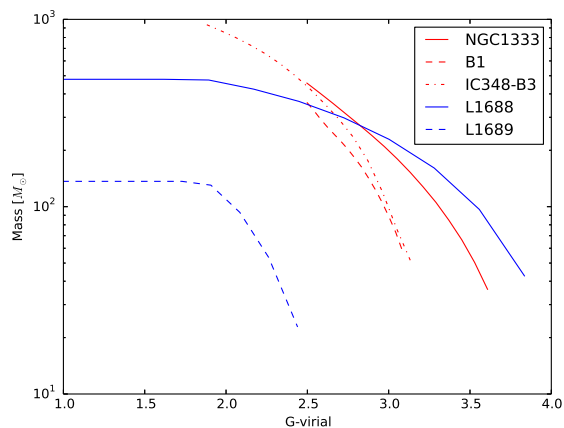}
\caption{\label{fig:massgvirial}The amount of molecular gas enclosed in regions
with different thresholds of the  \texttt{G-virial}. The $x$-axis is
the threshold of  \texttt{G-virial} which we take to define the region, and the
$y$-axis is the amount of gas that is enclosed in the region. }
\end{figure}

\begin{figure}
 \includegraphics[width=0.45\textwidth]{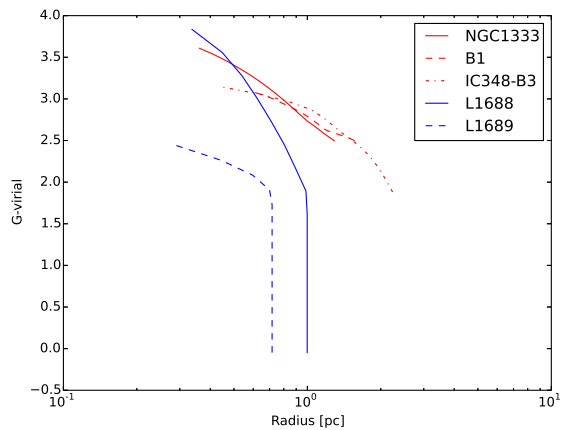}
\caption{\label{fig:gvirialradius} \texttt{G-virial} as a function of
radius for different regions. }
\end{figure}

We assess the importance of gravity in these regions with the  \texttt{G-virial}
method. In our method, a larger
 \texttt{G-virial} is related to a larger importance of gravity, and as
a result the importance of gravity at a given region can be quantified by
measuring the amount of gas at different thresholds of the  \texttt{G-virial}
parameter. In Fig. \ref{fig:massgvirial}, we plot the dependence of gas mass on
the  \texttt{G-virial} threshold for different regions, and in Fig.
\ref{fig:gvirialradius} we plot the dependence of the  \texttt{G-virial} on the radius. For all the
 regions, the  \texttt{G-virial} increases towards the centers. For
cluster-bearing regions such as NGC1333 in Perseus and L1688 in Ophiuchus, much
higher  \texttt{G-virial} values have been reached at their centers. 
This implies that gravity is more important for the clustered mode of star
formation.

\begin{figure}
\includegraphics[width=0.45\textwidth]{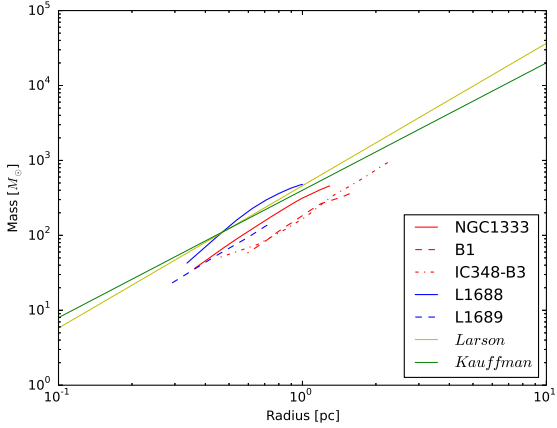}
\caption{\label{fig:massradius}Enclosed mass as a function of
radius for different regions. The scaling relation from
\citet{1981MNRAS.194..809L} and \citet{2010ApJ...716..433K} are added. For all
the curves, the
 \texttt{G-virial} decreases with increasing radii.  }
\end{figure}

With our method it is straightforward to derive relations such as the
mass-radius and velocity-size relation. Recently, there has been growing
interest in quantifying cloud structure in terms of various mass-size relations
\citep[e.g.,][]{2010ApJ...712.1137K,2010ApJ...716..433K}. In Fig. \ref{fig:massradius} we
plot the mass-size relation obtained using our  \texttt{G-virial} method (see
Section \ref{sec:justification}). 
Our mass-size relation is different from that of \citet{2010ApJ...716..433K} since
we use the \texttt{G-virial} contours to define our regions, and
mass is evaluated within a given region where the
 \texttt{G-virial} is larger than the threshold. In \citet{2010ApJ...716..433K}
the mass in evaluated within a region where the column density is larger than a
given threshold. In our case, along one single line of sight, only the gas that
stays within a given surface of a constant value of the  \texttt{G-virial} is
taken into account while in \citet{2010ApJ...716..433K} all the mass along the line of sight is taken into account.
 Similar to \citet{2010ApJ...716..433K}, we also found that cluster-bearing regions
such as L1688 and NGC1333 are more massive in terms of molecular gas than
the regions devoid of clusters at a give radius. In Fig. \ref{fig:massradius} we also plot the
scaling relations proposed in \citet[][$m(r)=460 M_{\odot} (r/{\rm pc})^{1.9}$]{1981MNRAS.194..809L},  and
\citet[][$m(r)=400 M_{\odot}(r/{\rm pc})^{1.7}$]{2010ApJ...716..433K} , and found that both provide
approximate descriptions of the structure of the regions.

\begin{figure}
\includegraphics[width=0.45\textwidth]{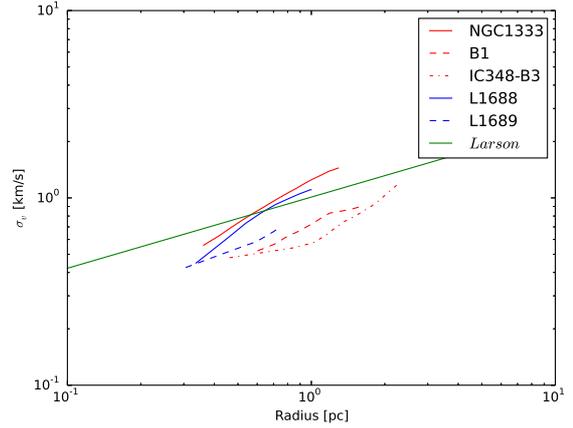}
\caption{\label{fig:velocitylinewidth}3D velocity dispersion as a function of
radius for different regions. The velocity-linewidth relation from
\citet{1981MNRAS.194..809L} is also plotted.  }
\end{figure}

 Molecular clouds are believed to be
turbulence-dominated, and this is mainly inferred from the fact that the clouds obey the velocity-linewidth relation
\citep{1981MNRAS.194..809L}. In Fig. \ref{fig:velocitylinewidth} we plot the 3D
velocity dispersion of our regions as a function of radius. Here the 3D velocity
dispersion is $\sqrt{3}$ times  the velocity dispersion evaluated in Eq.
\ref{eq:sigmav}. The scaling relation from \citet{1981MNRAS.194..809L} is also
plotted ($L/{\rm pc}\sim 1.01\times (\sigma_{\rm v}/ \;\rm km\;s^{-1})^{0.38}$).
In our case, the velocity-linewidth relation is evaluated for centrally condensed objects, and
a smaller physical scale is related to the inner part of a gas
condensation, whereas in \citet{1981MNRAS.194..809L} a smaller physical scale is
related to a smaller size of a subregion in a cloud \footnote{The physical
scale here can have different meanings; see
\citet[e.g.,][]{2004ApJ...616..943L,2004ApJ...615L..45H,2011ApJ...740..120R}.}.
For our centrally condensed objects, the scaling relation from
\citet{1981MNRAS.194..809L} seems to be valid. We also note that
cluster-bearing regions such as NGC1333 and L1688 have much larger velocity
dispersions  at large radii. 

\begin{figure}
\includegraphics[width=0.45\textwidth]{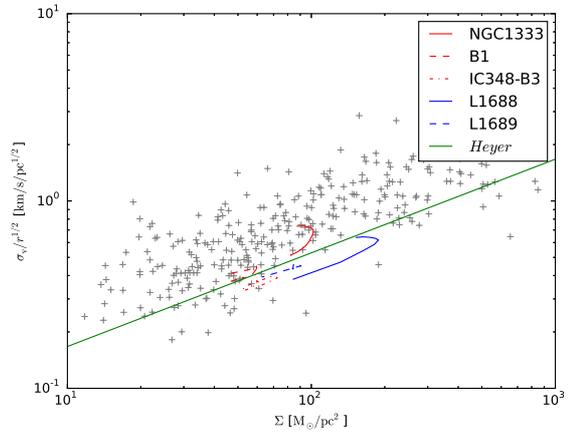}
\caption{\label{fig:heyer}  $\sigma_{\rm v}/r^{1/2}$ as a function of column density
$\sigma$ for the regions. In order to be consistent with
\citet{2009ApJ...699.1092H}, $\sigma_{\rm v}$ is the 1D velocity dispersion as
defined in Eq. \ref{eq:sigmav} and $\Sigma\equiv M/\pi r^2$
where $M$ is the mass $r$ is the radius. The solid straight line shows the
boundary below which the structures are gravitationally bound. It is defined as
$\sigma_{\rm v}/r^{1/2}=(\pi G/5 )^{1/2}\Sigma^{1/2}$. The gray crosses
come from the catalogue of giant molecular clouds studied in
\citet{2010ApJ...723..492R}. }
\end{figure}

\citet{2009ApJ...699.1092H} found a dependence of the scaling coefficient
$\sigma_{\rm v}/r^{1/2}$ of  Larson's relation  on the column
density, and attributed this to the fact that the clouds are in self-gravitational equilibrium. We plot
$\sigma_{\rm v}/r^{1/2}$ as a function of column density $\Sigma\equiv M/\pi
r^2$, where $M$ is the mass and $r$ is the radius of our regions (Fig.
\ref{fig:heyer}).
Since we decompose one region into a set of nested subregions based on the  \texttt{G-virial}, one single  region will appear as a
curve in Fig. \ref{fig:heyer}.
All our
regions are quite close to being gravitationally bound, and the scatter around the $y$-axis is much smaller than
what is found for Milky Way molecular clouds as studied in
\citet{2009ApJ...699.1092H}.
This difference arises mainly from the fact that the  \texttt{G-virial} method
tends to identify regions that are gravitationally coherent.

\section{Conclusion}\label{sec:conclu}
\subsection{Results and perspectives}
In this paper we proposed a {\it \emph{general method}} ( \texttt{G-virial}) for
studying the  structure molecular clouds. 
 Different from
the previous methods which decompose molecular clouds based on contours in the 2D plane or 3D
PPV data cubes, in this method a map is generated by taking all the
gravitational interactions between all the voxels in the 3D data cubes into
account.

The generated 3D  \texttt{G-virial} maps have a dimensionless unit that is the
same as that of the virial parameter, and a larger  \texttt{G-virial} is related
to a larger chance of being gravitationally bound. 
 Therefore, the method provides a global picture of gravity in the PPV
space. Using a hydrodynamical
simulation model and a simple spherical-symmetric model, we demonstrate the
connection between  \texttt{G-virial} calculated in the observed PPV data cube and the
 \texttt{G-virial} calculated from the model. We found that the two are
positively related. A larger  \texttt{G-virial} is therefore linked to a larger
chance of being gravitationally bound. Different from the virial parameter
which quantifies self-gravity, the  \texttt{G-virial} quantifies the effect of {\it \emph{global}} gravity on the gas where all the
interactions between all the particles in the 3D data cubes have been taken into
account.

A map of this kind enables us to to identify regions in molecular clouds in terms of
gravitationally coherent regions. Compared with previous ways of defining
regions based on emission, the regions identified from the  \texttt{G-virial}
maps more coherent and are less dependent on the technical parameters. Another
advantage of such a definition is that the regions identified are coherent
under gravity and are likely to collapse on their own. 

We also demonstrated how to use the  \texttt{G-virial} maps to quantify the structures of the
identified gas condensations in terms of mass-radius and velocity
dispersion-radius relations. 
We found that both the mass-radius relation
and the velocity-dispersion relation of the model can be reconstructed with the
help of the  \texttt{G-virial} with a good fidelity when the model is symmetric.
If the model is not symmetric, the  \texttt{G-virial} map can used to quantify
the structures of the gas condensations, and it allows us to compare different
structures.  

As examples, we analyzed the $^{13}$CO(1-0) emission from the Perseus and
Ophiuchus molecular clouds, and found that both can be decomposed into several regions which
are gravitationally coherent (Fig. \ref{fig:regions}). Moreover, the
cluster-bearing regions show higher values of  \texttt{G-virial} at the centers which implies
that gravity plays a more prominent role in these parts (Fig. \ref{fig:gvirialradius}).
We carried out an analysis of a total of five
regions identified in the Perseus and Ophiuchus molecular cloud, and derived mass-size
relations and velocity dispersion-size relations for the regions. We also found
that cluster-bearing regions are more massive at a given radius than those which
do not bear a cluster (Fig. \ref{fig:massradius}), and the cluster-bearing regions have a
higher velocity dispersion at the outer parts (Fig. \ref{fig:velocitylinewidth}). 

 The method is general and can be applied to a variety of objects observed in
 3D PPV space where gravity is supposed to play a role, and from the observations
 the method offers ways to quantify and compare these structures. We leave these
 possibilities for further explorations.  

 The importance of gravity in molecular clouds remains unclear. The effect of
 gravity as compared to kinetic motion is usually quantified using the
 virial parameter. The major uncertainty of the virial parameter comes
 from the definition of boundaries the regions. With the  \texttt{G-virial}
 method, we can eliminate this uncertainty by focusing on the regions that are gravitationally coherent.
 As shown in Fig. \ref{fig:heyer}, the G-virial-defined regions are much closer
 to being gravitationally bound than the giant molecular clouds studied in
 \citet{2009ApJ...699.1092H}. A study of a larger sample of molecular clouds can
 potentially tell if gravity is important throughout the Milky Way molecular
 clouds, and we leave this for a further study.
\subsection{Caveats}
Like all the other methods that quantify the ISM structures, our method is
influenced by the line-of-sight confusion. One advantage of our method is that
the line-of-sight confusion can be reduced by taking the velocity information
into account. 
For a single object, if the velocity dispersion increases with radius, the
 \texttt{G-virial} method can take this into account and reduce the
line-of-sight confusion effect. If different objects have different velocities,
they will be separated easily. If different objects with comparable mass
velocity are found on the same line of sight, the  \texttt{G-virial} method becomes inaccurate.

The  \texttt{G-virial} is a relative measure of the gravitational boundedness. As
a result, it is not possible to tell whether a region is gravitationally bound 
based on the  \texttt{G-virial}. However, a larger  \texttt{G-virial} value means
a region is more likely to be gravitationally bound. It is also
possible to derive the virial parameter afterwards based on the
iso-G-virial contours. \\
 
\noindent {\bf Code Bibliography}\\
astropy \url{http://www.astropy.org/}\\
astrodendro \url{https://dendrograms.readthedocs.org/en/latest/}\\
G-virial \url{http://gxli.github.io/G-virial/}\\
yt \url{http://yt-project.org/}

\begin{acknowledgements}
Guang-Xing Li is supported for this research through a stipend from the
International Max Planck Research School (IMPRS) for Astronomy and Astrophysics
at the Universities of Bonn and Cologne. Guang-Xing Li thanks Dr. Ke-Jia Lee for
helping with the calculations and thanks Dr. Arnaud Belloche for discussions. We
thank Eve Ostriker and Hao Gong for email exchanges.
This study makes use of data from the
COMPLETE survey. We also thank the \texttt{yt} and \texttt{astrodendro} teams
for making their codes available, and would like to thank Thomas Robitaille for
email exchanges. Dr. James Urquhart and Rosie Chen are acknowledged for
careful readings of the draft and for their helpful comments. We thank the
anonymous referee for several thorough and careful reviews
of the paper and his/her insightful comments, and thank Malcolm Walmsley for his
efforts.
\\

\end{acknowledgements}

\appendix
\section{Relation between  \texttt{G-virial} and the virial
parameter}\label{sec:gvirial:pvirial} The relative importance between
gravitational and kinetic energy is usually characterized with the virial parameter, which is introduced in
\citet{1992ApJ...395..140B}. To evaluate the virial parameter, it is necessary
to define a structure on which the virial parameter is calculated.

One difference between  \texttt{G-virial} and the virial parameter is that to
evaluate the
 \texttt{G-virial} no such boundary is needed. Therefore the  \texttt{G-virial}
can be viewed as a generalization of the virial parameter to a continuous
distribution of mass.

\begin{figure}
\includegraphics[width=0.5 \textwidth]{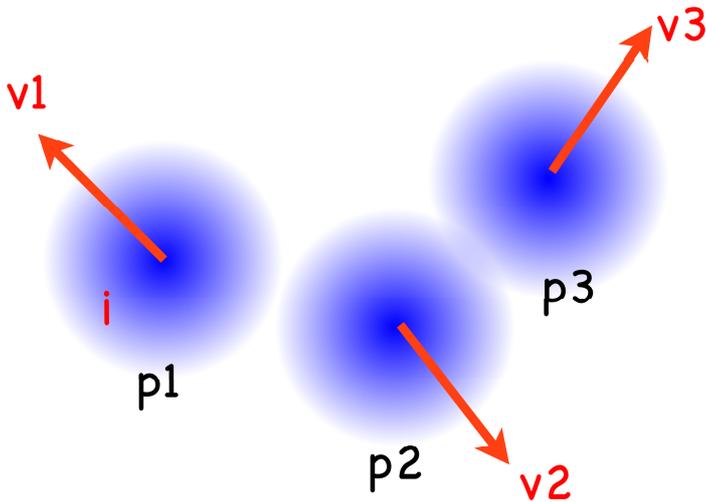}
\caption{\label{fig:3p}The difficulty of finding a proper
center of mass.
We consider three clumps p1, p2, p3 moving at velocities v1, v2, v3. If we are
interested in whether $i$ is gravitationally bound, since the majority of attraction
comes from p1, physically the center of mass should be the center of mass of p1
instead of the center of mass of the three clumps p1, p2, p3, since p2 and
p3 do not contribute much gravitational attraction to $i$ compared to p1.
Therefore to estimate the gravitational boundedness, a naive calculation of the
center of mass where all the mass in the whole cloud is included is not
appropriate.  In the case of a complicated distribution of gas, finding the
center of mass is not straightforward.  }
\end{figure}

To illustrate the physical meaning of the  \texttt{G-virial}, we define a
quantity called {\it \emph{particle virial, p-virial}}, which is the ratio between the
gravitational energy of a particle to its kinetic energy $\alpha_{\rm
p-virial}=E_{\rm p}/2 E_{\rm k}$. The potential energy $E_{\rm p}$ is
determined by $m\times \phi$ where m is the mass of the particle and $\phi$ is
the gravitation potential, and $E_{\rm k}$ is determined as $1/2\times m \;
(v-v_{\rm c})^2$.  To define the kinetic energy, the velocity of the center of
mass $v_{\rm c}$ is needed.
Therefore {\it \emph{p-virial}} is only suitable in the cases where a center of mass
can be easily found. In the case of a molecular cloud, this is not
straightforward  because inside a molecular cloud the condensations can
move at different velocities. The velocity of the center of mass depends on
which clumps are included in the calculation, which is not unique. 
This difficulty is illustrated in Fig. \ref{fig:3p}.

One major advantage of the  \texttt{G-virial} over the {\it p-virial} is that no
center of mass is needed. The  \texttt{G-virial} is the sum of the gravitational
boundedness, and to evaluate it we do not need to introduce the center of mass.
Therefore, the  \texttt{G-virial} can provide estimates of the importance of
gravity when the geometry of the object is complicated and the center of mass is
not well defined.

\section{Dependence on of  \texttt{G-virial} the smoothing velocity
$c_0$}\label{sec:c0} The only free parameter in our method is the smoothing velocity $c_0$ in Eq.
\ref{eq:gvirial2}. We added this parameter to avoid the
divergence of Eq.
\ref{eq:gvirial2} when the separation of velocity is zero. 

Physically, if the velocity separation is zero, and if two gas particles 
spatially coincide with each other, the gravitational interaction will cause
them to collide. However,  the velocity separation cannot be zero
since the gas also has velocity dispersions that have either thermal or
non-thermal origin \citep{1981MNRAS.194..809L}. A physically lower limit of
$c_0$ should be the sound speed.

In reality, this is also affected by the superposition along the line of
sight, since different gases that are physically unassociated can stay along the
same line of sight and therefore appear to be gravitationally bound.
Therefore, it is difficult to find a unique value of $c_0$ based on first
principles. In our case, we choose $c_0$ to be 1 km/s, which is larger than and
still comparable to the sound speed. 

\begin{figure}
\includegraphics[width=0.48\textwidth]{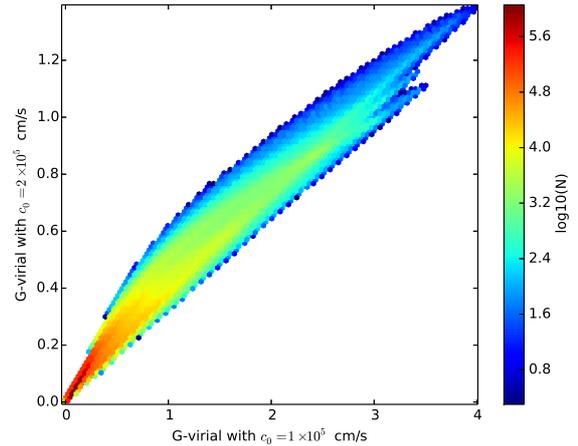}
\caption{\label{fig:compare} A comparison of  \texttt{G-virial} map of the
Ophiuchus molecular cloud under different $c_0$. The horizontal axis is the
 \texttt{G-virial} calculated assuming $c_0=1\;\rm km/s$ and the vertical axis is
the  \texttt{G-virial} calculated assuming $c_0=2\;\rm km/s$. The grayscale image
stands for the number of voxels that fall into each bin.  }
\end{figure}

Here  we investigate how our results can be affected by
 the parameter $c_0$. First, we made a  \texttt{G-virial} map by
assuming $c_0=1\;\rm km/s$ and then we made another map by assuming $c_0=2\;\rm
km/s$. Finally, we made a voxel-by-voxel comparison of the maps (Fig.
\ref{fig:compare}). We found  the major effect of changing $c_0$ on the
absolute values of  \texttt{G-virial}: if we change $c_0$ to $2\;\rm
km/s$ the  \texttt{G-virial} decreases by a factor of $\sim 2$. However, there is good
correspondence between the old and new  \texttt{G-virial} values. Therefore,
changing $c_0$ affects the absolute values of the  \texttt{G-virial};  however, the
relative values are unaffected. Therefore, all our
figures should be unaffected by a change of $c_0$ and the  \texttt{G-virial} axes
of Fig.
\ref{fig:massgvirial} and Fig.
\ref{fig:gvirialradius} will change accordingly. 

In the case where the  \texttt{G-virial} is applied to different data cubes, in
order for the results to be comparable, a unique choice of $c_{\rm 0}$ is
necessary.

\section{3D renderings of $ \texttt{G-virial}$ of the Perseus and Ophiuchus
molecular clouds}
 \label{sec:rendering} In order to provide intuitive representations of
our
 \texttt{G-virial} maps, we present volume renderings of the 3D PPV data cubes with \texttt{yt}
\citep{2011ApJS..192....9T} in Fig. \ref{fig:channel}. {\bf A movie can be
found in \url{https://vimeo.com/115595991}.}

\begin{figure}
\includegraphics[width=0.48\textwidth]{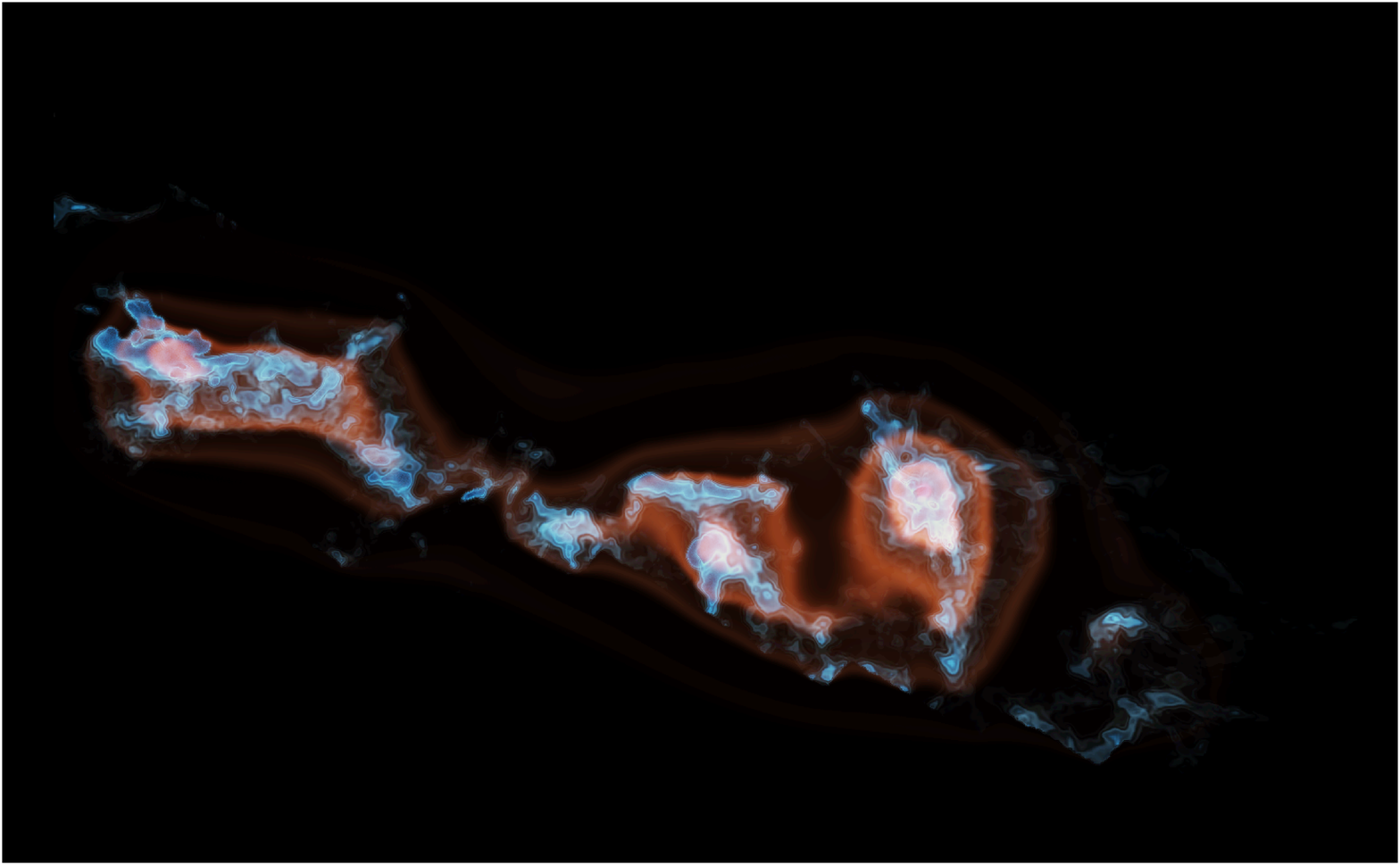}

\includegraphics[width=0.48\textwidth]{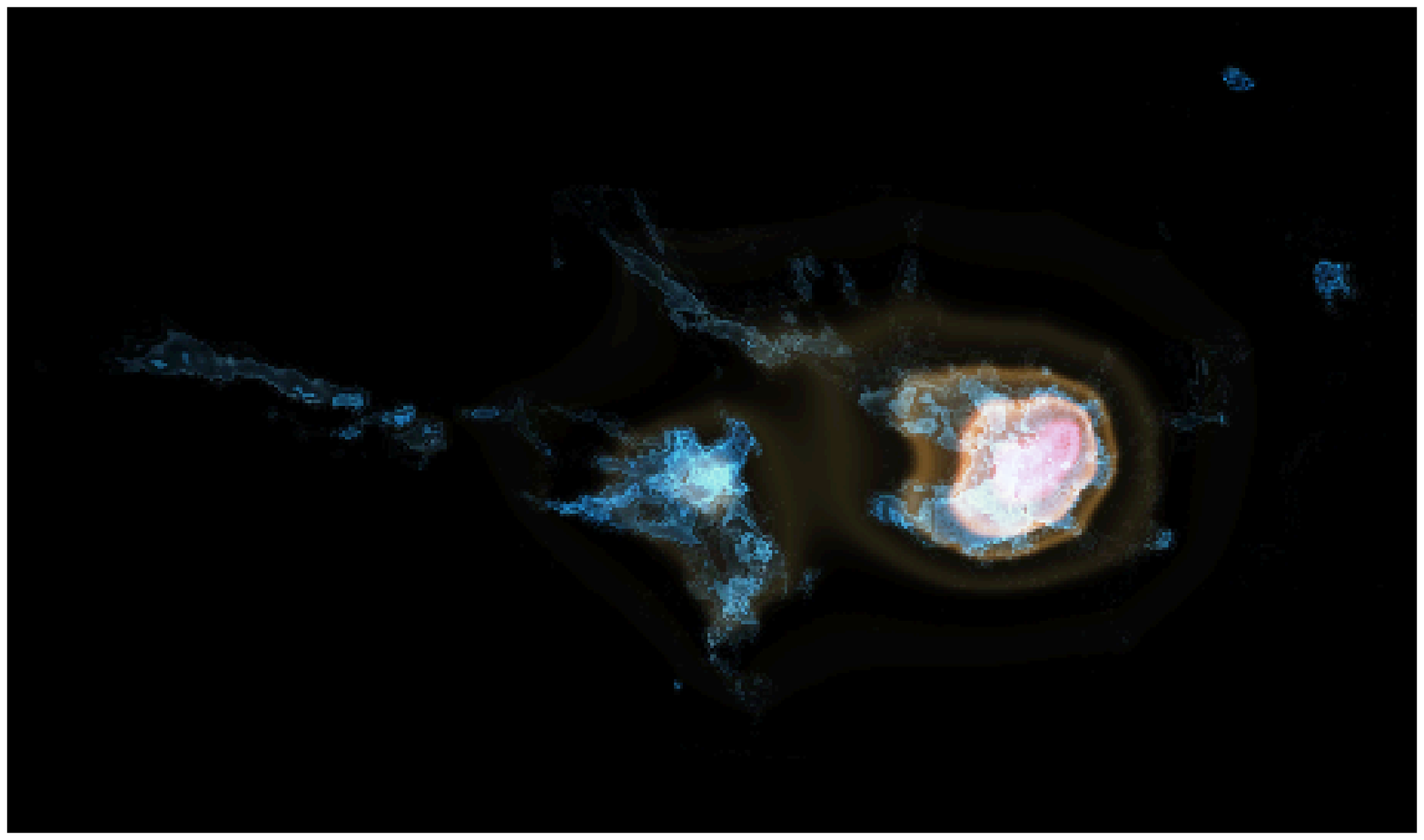}
\caption{\label{fig:rendering} Volume rending representations of the
$^{13}$CO(1-0) emission and the corresponding  \texttt{G-virial} map. The upper
panel shows the result from the Perseus molecular cloud and the lower panel
shows the result from the Ophiuchus molecular cloud. Blue  stands for
$^{13}$CO(1-0) emission and  red and orange  stand for 
 \texttt{G-virial}. Both clouds are projected along the velocity
direction. For the Perseus molecular cloud, the iso-surfaces of
$^{13}$CO(1-0) emission start from 0.3 K and increase in  steps of 0.64 K. The
contours of the  \texttt{G-virial} starts from 1.2 and increase in  steps of 0.8. 
For the Ophiuchus molecular cloud, the contours of $^{13}$CO(1-0)
emission start from 0.3 K and increase in  steps of 1.24 K. The
contours of the  \texttt{G-virial} starts from 1.2 and increase in  steps of 0.8.
}
\end{figure}

\section{Comparison with other methods}\label{sec:appendix:comparision}

The previous methods to quantify the structure of molecular condensations such
as \texttt{Clumpfind} and  \texttt{Dendrogram} focus on the structure
traced by the intensity map. The {\it \emph{Clumpfind}} method tends to produces isolated
structures. The {\it \emph{Dendrogram}} produces hierarchical
representations of the nested isosurfaces in 3D molecular line
data cubes, and offers multi-scale decompositions. It can be applied to a much
larger variety of situations.
Both methods work in  2D (position-position) and 3D (PPV) space and both methods
decompose the data to some extent. 

The  \texttt{G-virial} method presented in this paper is also based on a map of
intensity in 3D PPV space. However, its output is neither a list of
structures (as in the case of the \texttt{Clumpfind}) nor a tree representation
of a hierarchy of structures (as in the gas of the \texttt{Dendrogram}).
Instead, it produces a map of the importance of gravity in the 3D PPV. The output is also a map  3D PPV space. The
 \texttt{G-virial} map itself does not provide a decomposition of the data.
To decompose the cloud into regions and to analyze their properties, other
methods are needed. 

In this work, we use the \texttt{Dendrogram} to identify
gravitationally coherent regions from the  \texttt{G-virial} map, and use the iso-G-virial contours to
quantify the properties of the regions in the $M$-$r$ and $\sigma_{\rm v}$-$r$
plane. The  \texttt{G-virial} method is not a replacement of the other methods,
but it provides a new map on which those methods could be applied.

\bibliography{paper_gravity}
\end{document}